\documentclass[sigconf]{acmart}

\usepackage{graphicx}
\graphicspath{{./images/}}
\usepackage{booktabs} % for formal tables
\usepackage{multirow}
\usepackage{amsmath}  % for aligning equations
\usepackage{subcaption}
\usepackage{caption}
\usepackage{multirow}
\usepackage{multicol}
\usepackage{hyperref}
\usepackage{url}
\usepackage{balance}
\usepackage{verbatim}
\usepackage{xspace}
\usepackage{mathtools}
\usepackage{comment}
\usepackage{enumitem}

\newcommand{\elixir}{\textsc{Elixir}\xspace}
\newcommand{\prince}{\textsc{Prince}\xspace}
\newcommand{\recwalk}{\textsc{RecWalk}\xspace}

\theoremstyle{definition}

\copyrightyear{2021}
\acmYear{2021}
\acmConference[WWW '21]{Proceedings of the Web Conference 2021}{April 19--23,
2021}{Ljubljana, Slovenia}
\acmBooktitle{Proceedings of the Web Conference 2021 (WWW '21), April 19--23, 2021,
Ljubljana, Slovenia}
\acmPrice{}
\acmDOI{10.1145/3442381.3449848}
\acmISBN{978-1-4503-8312-7/21/04}

\begin{document}
	
	\title{ELIXIR: Learning from User Feedback on Explanations\\to Improve Recommender Models}
	% \title{Actionable Explanations: Learning from User Feedback on\\
	%	Item Pairs to Improve Recommender Models}
	% \title{Actionable Explanations: Learning from User Feedback on Item Pairs to Improve Recommender Models}
	% \title{ELIXIR: Efficient Learning from Item-level Explanations in Recommendation Systems}
	
	\author{Azin Ghazimatin}
	%\authornote{Dr.~Trovato insisted his name be first.}
	\affiliation{%
		\institution{Max Planck Institute for Informatics, Germany}
		% \\ Saarland Informatics Campus, Germany}
	}
	\email{aghazima@mpi-inf.mpg.de}
	
	\author{Soumajit Pramanik}
	\authornote{This work was done while the author was at the MPI for
		Informatics.}
	\affiliation{%
		\institution{IIT Bhilai, India}
		% \\		Palaiseau, France}	
	}
	\email{soumajit@iitbhilai.ac.in}
	
	\author{Rishiraj Saha Roy}
	\affiliation{%
		\institution{Max Planck Institute for Informatics, Germany}
		% \\ Saarland Informatics Campus, Germany}
		%\streetaddress{Address}
		%\city{City}
		%\state{Country}
	}
	\email{rishiraj@mpi-inf.mpg.de}
	
	\author{Gerhard Weikum}
	\affiliation{%
		\institution{Max Planck Institute for Informatics, Germany}
		% \\ Saarland Informatics Campus, Germany}
	}
	\email{weikum@mpi-inf.mpg.de}
	
	{\tiny }\renewcommand{\shortauthors}{Ghazimatin et al.}
	
	\newcommand\BibTeX{B{\sc ib}\TeX}
	
	\newcommand{\squishlist}{
		\begin{list}{$\bullet$}
			{ \setlength{\itemsep}{0pt}
				\setlength{\parsep}{3pt}
				\setlength{\topsep}{3pt}
				\setlength{\partopsep}{0pt}
				\setlength{\leftmargin}{1.5em}
				\setlength{\labelwidth}{1em}
				\setlength{\labelsep}{0.5em} } }
		
		\newcommand{\squishend}{
	\end{list}  }
	
	% !TeX root = ../2021-www-fp-elixir.tex
\begin{abstract}

System-provided explanations for recommendations are an important component towards transparent and trustworthy AI. In state-of-the-art research, this is a one-way signal, though, to improve user acceptance. In this paper, we turn the role of explanations around and 
investigate how they can contribute to enhancing the quality of the generated recommendations themselves. We devise
a human-in-the-loop  
framework, called \elixir, where user feedback on explanations is leveraged for pairwise learning of user preferences.
\elixir
leverages
feedback on pairs of recommendations and explanations to learn user-specific latent preference
vectors, overcoming 
sparseness by label propagation with item-similarity-based neighborhoods. Our framework is instantiated
using generalized graph recommendation 
via Random Walk with Restart.
Insightful experiments with a real user study
show significant improvements in
movie and book 
recommendations over item-level feedback.
\end{abstract}

	
	
% 	\keywords{Transparency, Explanations, User feedback, Recommender Systems}
	
	\maketitle
	
	% !TeX root = ../2021-www-fp-elixir.tex
\section{Introduction}
\label{sec:intro}

\textbf{Motivation.} 
%
%GW: one par on scrutable recommendations and actionable explanations
%
Generating explanations for recommendations like 
%movies, books, and news, 
movies, music or news,
by online service providers, has gained 
% high attention in research and industry
high attention in academic and industrial research
%a very high research 
%traction in recent years
%~\cite%{ghazimatin2020prince,balog2019transparent,balog2020measuring,tintarev2012evaluating}.
%%%GW: all these references are important for the related work section, but the intro should be sparser on references
%must cite the survey article here, and one high-caliber paper from industry
%(or an industrial keynote, could cite Lalmas or some Google bigshot here if their talks contain large parts on explanations?)
%rishi: both the balog papers are from google research, so I guess
% one of them would fit best here. also added a microsoft research paper
\cite{zhang2020explainable,balog2020measuring,zhao2019personalized}.
While early methods generated simple explanations such as
``you liked $x$ and other users also liked $x$ and $y$ ...'', modern approaches have become much more
sophisticated (see, e.g.,~\cite{zhang2020ears}).
%Hand-in-hand,
%explainable recommendation~\cite{ma2019jointly,wang2019explainable,ai2018learning} %(see~\cite{zhang2020explainable} for a review), where a key goal of the model designer %is to facilitate discovery of faithful or causal explanations for recommended items, %has almost developed a community of itself~\cite{zhang2020ears}.
%%%GW: above is unnecessary paraphrasing/ramification
%
%GW: no raise the points of scrutability/actionability
%
A key goal is to enhance user trust by transparent and scrutable recommendations, so that users understand how the recommended item relates to
their prior online behavior (search, clicks, likes, ratings, etc.) (see, e.g., % \cite{balog2019transparent,balog2020measuring}).
% actually fairy fits exactly here
% balog2020 does not really fit... (regarding the relationship)
\cite{balog2019transparent,ghazimatin2019fairy}).
Moreover, it is desirable that explanations are {\em causal} and {\em actionable},
meaning that i) they refer only to the user's own action history and not to potentially
privacy-sensitive cues about other users
(see, e.g.,~\cite{ghazimatin2020prince}) 
and ii) the user can act on the explanation items by giving confirmation or 
refutation signals that affect future recommendations. 
Critique-enabled
recommendation models~\cite{luo2020deep,chen2012critiquing,jin2019musicbot, lee2020explanation} pursue these goals,
but are restricted to user feedback on the recommended items
and associated content (e.g., text snippets from item reviews),
disregarding the \textit{explanation items}.
In this paper, we extend this regime of actionable user feedback to 
the explanations themselves, by obtaining additional cues from users in a lightweight manner and incorporating them into
an active learning framework
% enhanced learning
to improve future recommendations.

\begin{figure} [t]
	\centering
	\includegraphics[width=\columnwidth]{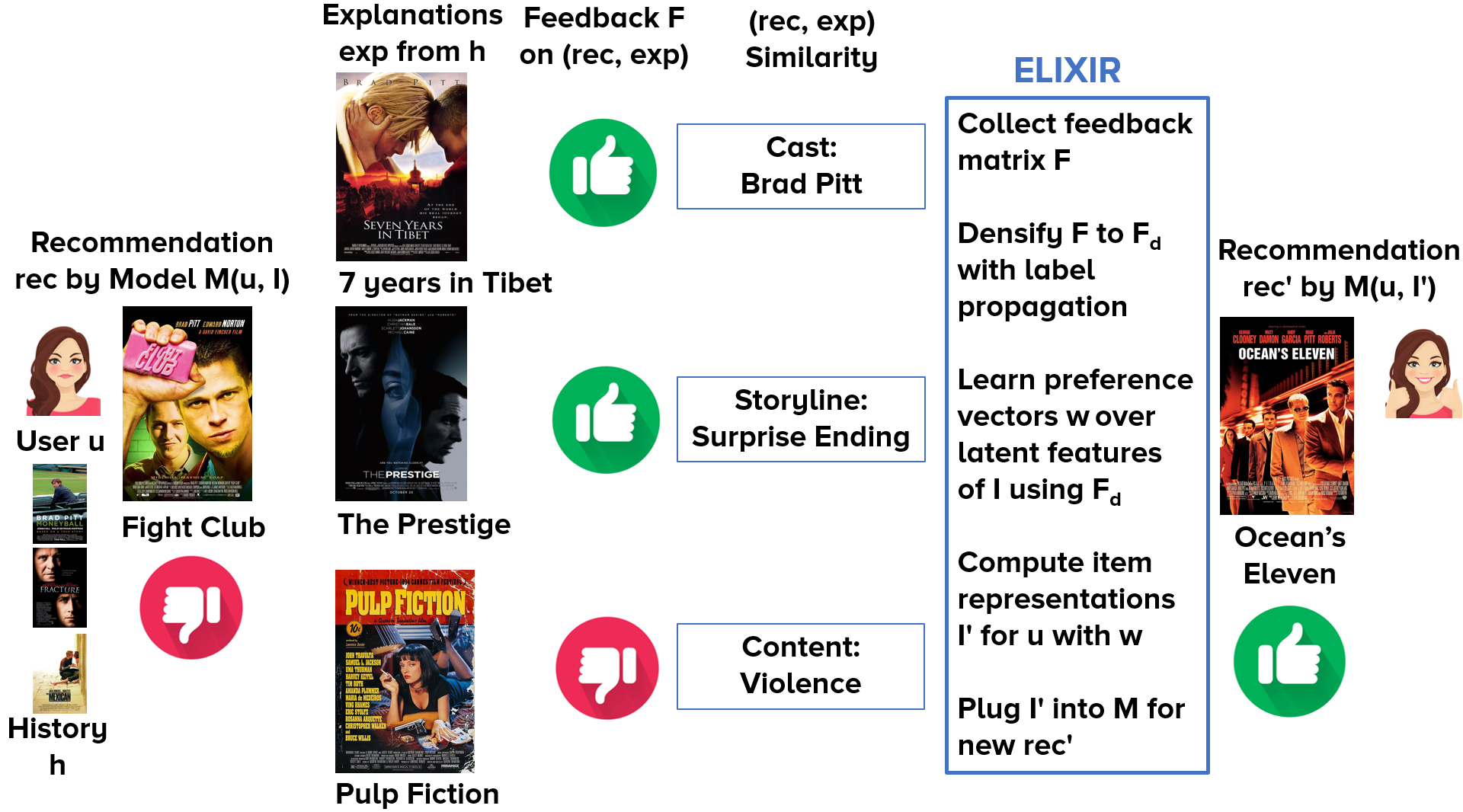}
%	\vspace*{-0.7cm}
	\caption{Example illustrating the intuition for \elixir.}
	\label{fig:elixir}
	% \vspace*{-1cm}
\end{figure}
% \GW{drop figure, textual description should be clear enough}

 \vspace*{0.1cm}
\noindent\textbf{Example.}
Fig.~\ref{fig:elixir} shows an illustrative scenario.
%%%GW: do we need the figure at all??? the text says everything 
%%% by reformating the text we can also give it visual highlighting (e.g., itemize the three exp items)
Alice receives a recommendation for the movie 
%\textit{Once Upon a Time in Hollywood} 
\textit{Fight Club}
($rec$) based on her online history and factors like item-item similarities.
This is accompanied by an explanation referring to three items, all previously liked by Alice
and being similar, by some aspects, to $rec$.
%\squishlist
%\item 
We have $exp1$: \textit{Seven Years in Tibet},
%(previously liked by Alice), 
%\item 
$exp2$: \textit{The Prestige}, and
%(similar in some aspects to $rec$) and
%\item $exp3:$ \textit{Inglorious Basterds} (having the same director as $rec$).
%notice that Brad Pitt also an actor in this one
%\item 
$exp3$: \textit{Pulp Fiction}.
%(having the same director as $rec$).
%\squishend
%\noindent 
The system generated these three items for explanation because:
\squishlist
\item $exp1$ features the actor Brad Pitt who also stars in $rec$,
\item $exp2$ has a surprise ending, similar to $rec$,
% and
\item $exp3$ contains violent content, like $rec$.
\squishend

% Now suppose that Alice loves Brad Pitt and twisted plots but hates 
% extremely violent movies.
Now suppose that Alice loves Brad Pitt and surprise endings but hates
disturbing violence (she likes \textit{Pulp Fiction} for other reasons like
its star cast and dark comedy, that dominated her opinion, despite
the violence).
When receiving $rec$ with the above explanation,
Alice could give different kinds of feedback. 
The established way is to simply dislike $rec$, as a signal from
which future recommendations can learn. However, this would completely
miss the opportunity of learning from how Alice views the three
explanation items. Alice could also dislike the explanation as whole, but this
would give only a coarse signal, too, and would appear conflicting
with the fact that she previously liked $exp1, exp2$ and $exp3$,
confusing the recommender system. 
The best feedback would be if Alice could
inform the system that she likes Brad Pitt and surprise endings
but dislikes violence, for example,
by checking item properties or filling in a form or questionnaire.
However, this would be a tedious effort that few users would engage in.
Also, the system would have to come up with a very fine-grained feature space of properties,
way beyond the usual categories of, say, movie genres. 

\vspace*{0.1cm}
\noindent\textbf{Problem Statement.}
The goal in this paper is to leverage user feedback
% from situations.
on explanations.
This entails two major problems:
\squishlist
\item \textit{Feedback:} How can we elicit user feedback on properties of both recommended and explanation items
in a lightweight manner, without burdening the user with too much effort?
\item \textit{Actionability:} If we are able to obtain such refined feedback, how can the recommender system learn from it to improve its future outputs
for the user?
\squishend

\vspace*{0.1cm}
\noindent\textbf{Approach.} 
%The clear inference from the above chain of reasoning is that the system should instead try to unearth what about the explanations the user liked and disliked, \textit{in the context} of the recommended item. Such fine-grained preferences should then be incorporated into the underlying model to produce more informed recommendations. This is precisely the thesis of this paper, a framework coined \elixir: \underline{\textbf{E}}fficient \underline{\textbf{L}}earning from \underline{\textbf{I}}tem-level e\underline{\textbf{X}}planations \underline{\textbf{I}}n \underline{\textbf{R}}ecommenders.
%
This paper presents \elixir
% need to introduce the full form now
(\underline{\textbf{E}}fficient \underline{\textbf{L}}earning from \underline{\textbf{I}}tem-level e\underline{\textbf{X}}planations \underline{\textbf{I}}n \underline{\textbf{R}}ecommenders),
a novel framework, for leveraging explanations
% in order
to improve future recommendations. 

We address the \textit{Feedback} problem by asking users for a binary like/dislike signal about the
% relatedness
similarity 
of an explanation item $exp$ to the recommended item $rec$. This can be thought of as assessing the
quality of the {\em item pair} $\langle rec, exp \rangle$, but it is important that one of these is an
explanation item that was faithfully produced by the recommender system specifically to justify $rec$.
Our experiments compare this choice against asking for
% relatedness assessments
assessments on the similarity of
% of randomly sampled item pairs
the recommendation with the least relevant items in the user's profile,
which turns out to be inferior. % check results later
As we consider only causal explanations that refer to the user's own history of actions, the user should be reasonably familiar with item $exp$.
This kind of feedback is more refined than simply disliking the entire recommendation.
The feedback is optional and can be given for any subset of the possible $\langle rec, exp \rangle$ pairs.
Most importantly, the user is not burdened with identifying relevant properties 
%or categories 
of items,
to further explain her feedback to the system.
So \elixir builds on very lightweight
and proactive
user feedback.

We address the \textit{Actionability} problem by extending state-of-the-art recommender models
with a user-feedback matrix that encodes the like/dislike signals on $\langle rec, exp \rangle$ pairs.
Since this matrix is inevitably sparse, \elixir densifies
% and regularizes
this input by means of
label propagation on item neighborhoods \cite{xiaojin2002learning}.
%sparse due to the huge number of potential item pairings, \elixir then ``densifies'' it using label propagation~\cite{xiaojin2002learning} on a neighborhood
%built by locality sensitive hashing (LSH). 
%GW: LSH is only important here (in the intro) if efficiency were a key concern
%
%The cost for constructing complete neighborhoods from the similarity of item-pairs
%($10^3$ items lead to a total of $10^6$ pairs, and hence $10^{12}$ possible similarity calculations)
%is simply exorbitant.
%Thus, 
%%%GW: techniques like LSH as sim pre-filter are very standard, I would not claim this as a novel contribution (and efficiency is not considered in the experiments) - so this needs to be toned down and kept short
To avoid the huge cost of computing similarities for all item pairs, 
we employ locality sensitive hashing (LSH) to find the closest items to every $\langle rec, exp \rangle$
tuple, thereby making \elixir efficient and tractable.

The core of our method is the learning of user-specific latent vectors that capture user preferences,
%learning user-specific preference vectors over latent item features using
by combining the densified feedback matrix and a prior item-item similarity matrix
through regularized optimization that models the signals in the feedback matrix as soft constraints.
The latent vectors would reflect that Alice 
%pays more positive attention to 
%actors like Brad Pitt than to directors like Tarantino
loves Brad Pitt and sophisticated plots but dislikes violent movies
% the actors in movies
% (she likes Brad Pitt) than to the directors (she dislikes Tarantino)
-- without referring to these properties, 
all by means of learning latent representations from lightweight feedback.
The per-user vectors are plugged into the actual recommender system to learn
user-specific item representations
% that enhance
for future recommendations.

We instantiate the \elixir framework
% for two popular families of recommenders:
in a popular family of graph-based recommenders:
Random Walk with Restart (RWR) (see, e.g., \cite{nikolakopoulos2019recwalk}),
% and 
% Matrix Factorization (MF) (see, e.g., \cite{koren2009matrix,nikolakopoulos2019eigenrec}),
from which
explanations 
%are generated using intuitive techniques.
can be generated in a faithful and causal manner
% (see, e.g.,~\cite{ghazimatin2020prince,cheng2019incorporating}).
% rishi: the second reference is a bit tricky
(see, e.g.,~\cite{ghazimatin2020prince,wang2020learning}). 
% Our code is publicly available at \url{https://github.com/azinmatin/elixir/}.

%Significant improvements in movie (MovieLens) and book (GoodReads) recommendation experiments involving real users
%as well as extensive simulations, show the viability of our proposal.
%% \elixir takes explanations beyond their superficial use today,
%and makes them actionable by end users towards a direct improvement of individualized recommendation models.

\vspace*{0.1cm}
\noindent\textbf{Contributions.} The salient contributions of this paper are:
\squishlist
	\item \elixir is, to the best of our knowledge, 
%a framework that makes explanations concretely actionable by end users such that a direct improvement of individualized recommendation  is possible. Specifically, as far as we know, it is the first to learn from user feedback on system-generated explanations to improve recommender models.
the first framework that leverages user feedback on explanation items, thus making explanations actionable, whereas prior works only tapped into feedback on recommendation items.
	\item \elixir elicits lightweight user feedback to learn user-specific item representations, and
incorporates these into the recommender model, instantiated with the RWR methodology.
% and MF families.
%
%GW: the following is too long, too verbose, and too self-praising
%	\item We show that \elixir is more than a concept prototype: longitudinal studies with real users spanning six weeks show the tractability of \elixir for both models, in movie and book recommendation. In the course of this
%	study, the users went through the whole pipeline of generating real profiles from their preferences, interacting with the recommenders,
%	and evaluating both $rec$ items and $(rec, exp)$ pairs.
%	\item We publicly release the data collected during the user study conducted for evaluating \elixir. As far as we know, this is the first resource of its kind where we have end-to-end information on user profiles, %%recommendations, causal explanations, 
%	and item- and pair-level feedback.
%	In total we have about xxx item-level ratings and yyy pair-level ratings in each of the two setups. We believe that such a resource will be very valuable to the community to further explore feedback on explanations.
\item We report
% on
experiments with 
%both simulated user behavior derived from existing data collections
%and 
data from
a longitudinal user study in two domains: (i) movies, and (ii) books. 
%Both
The results
demonstrate the viability of
% the
\elixir
% method
and 
its substantial gains
in recommendation quality
% compared to baselines.
over item-level feedback.
\item The user study data and all code for this project are released as research resources
at 
%\url{https://bit.ly/2ZbbNRe} 
\url{https://www.mpi-inf.mpg.de/impact/learning-from-feedback-on-explanations}
and \url{https://github.com/azinmatin/elixir/}.
% , with explicit consent by the participating users and approval
% by our Ethics Review Board, upon publication.
\squishend

\section{The \elixir Framework}
\label{sec:framework}

%3.1 <Main idea>
%3.1.1 Label Propagation
%User-specific preference vector
%Transformation function
%Feedback collection, densification, incorporation
%Hard constraints to combined objective
%Regularization
%3.2 Instantiation in RWR
%3.3 Instantiation in MF
In this section, we describe the components of \elixir and present its 
instantiation for a state-of-the-art recommender, \recwalk~\cite{nikolakopoulos2019recwalk}, 
that is based on random walks with restart (RWR). 
Table~\ref{tab:notation} summarizes
%notions used in this section.\\
concepts and notation
% that will be
discussed in this section.

\begin{table} [t]
	\centering
	\resizebox{\columnwidth}{!}{
	\begin{tabular}{l l} \toprule
		\textbf{Notation}					& \textbf{Concept}										\\	\toprule
		$u$                    				& A single user                                         \\
		$v$                     			& A single item                                       	\\
		$\vec{v}$               			& Latent vector for item $v$                            \\
		$d$                     			& Number of latent dimensions                           \\
		$U$                     			& Set of all users                                      \\
		$I$                     			& Set of all items                                      \\
		$H_u$                   			& Interaction history of user $u$                       \\	
		$F_u(v_i, v_j)$ 					& Feedback on item pair $(v_i, v_j)$ by user $u$        \\	\midrule
		$v_{ij}$							& Pseudo-item for item pair $(v_i, v_j)$ in LP 			\\
		$W$									& Affinity matrix for LP								\\	
		$F_u^d$ 							& Densified feedback matrix for user $u$ after LP       \\	
		$m$									& Number of non-zero elements in $F_u^d$                \\ \midrule		
		$sim(\vec{v_i}, \vec{v_j})$			& Similarity of item pair $(v_i, v_j)$             		\\		
		$\vec{w_u}$                 		& Preference vector for user $u$                        \\
		$g(\vec{v}, \vec{w_u})$     		& Latent representation of item $v$ for user $u$        \\		
		$\gamma$              				& Regularization coefficient for learning $w_u$  		\\	\midrule
		$G$                     			& Graph on which the RWR recommender is run   			\\		
		$N$                     			& Set of nodes in graph                   				\\
		$E$                     			& Set of edges in graph                   				\\		
		$T_N, T_E$              			& Types of nodes and edges in graph               		\\		
		$\theta \; (\theta_N, \theta_E)$	& Mapping functions (of graph nodes and edges) to types \\				
		$A$                     			& Matrix of user-item interactions for RWR              \\
		$S$                     			& Item-item similarity matrix for RWR                   \\
		$\vec{e_u}$                         & One-hot vector for user $u$                           \\
		$\alpha$                			& Restart probability in RWR                            \\
		$\beta$                 			& Probability of walking over interaction edges in RWR  \\ 
		$rec$           					& Recommendation item                               	\\
		$exp$								& Explanation item										\\	\bottomrule
	\end{tabular}}
	\caption{Notation for salient concepts in \elixir.}
	\label{tab:notation}
	\vspace*{-0.7cm}
\end{table}

\subsection{Feedback collection}
\label{subsec:collect}
%\textbf{Feedback Collection.}

\elixir enables recommender systems to combine individual item ratings with feedback 
on pairs of recommendation and explanation items. The set of item-level signals 
$H_u$ refers 
to the set of individual items
%with which $u$ has already interacted. 
that appear in the interaction history of user $u$.
Denoting the universe 
of all items by $I=\{v_1, v_2, ...v_{|I|}\}$, 
we have $H_u \subseteq I$ and usually $|H_u| \ll |I|$.

While most recommenders train a
user model solely based on $H_u$,
\elixir exploits signals from
user feedback on item pairs from recommendations and their explanations. We denote this pair-level 
feedback by the matrix $F_u\in \{-1, 0, +1\}^{|I| \times |I|}$. %, where $m = |I|$. 
The matrix entry $F_u(v_i, v_j)$ represents user $u$'s feedback on 
recommendation item $v_i$ and 
explanation item $v_j$. To collect such feedback, we ask users whether they like/dislike the 
similarity between items $v_i$ and $v_j$. 
%\GW{should we really call this similarity? isn't more like relevance of exp for rec? that is, an asymmetric, kind of conditional judgement?}
We encode a user's liking, no feedback, and disliking with 
$+1$, $0$ and $-1$, respectively. 
%Note that matrix $\mathcal{F}$ becomes triangular if we assume that the items in the user's history will 
%never be chosen as a future recommendation, i.e., $v_i \in I \setminus H_u$. 
%\GW{why triangular? symmetry? not really? ???}

\subsection{Feedback densification}
\label{subsec:densify}
%\textbf{Feedback Densification.}

As the set of all item-pairs is very large, we expect matrix $F_u$ to be extremely sparse. 
To mitigate this sparseness, we use the label propagation (LP)
algorithm~\cite{xiaojin2002learning}
%by Zhu and Ghahramani
on a
% special
graph where nodes are \textit{pairs of items}, and
edges represent the \textit{similarity between item-pairs}. 
To define such a graph, we introduce the concept of
a pseudo-item $v_{ij}$ for each labeled pair of items $(v_i, v_j)$ 
(that models an item that is like a mixture of the properties of the two items in the pair)
where $F_u(v_i, v_j) \neq 0$, with $\otimes$ denoting the element-wise product:
\begin{equation}
	\vec{v_{ij}} = (\vec{v_i} \otimes \vec{v_j})^{\frac{1}{2}}
	\label{eq:pseudo-item}
\end{equation}
where $\vec{v_i}$ is the feature vector for item $v_i$. Depending upon
the recommender model, item 
features are either learned by the model~\cite{koren2009matrix,he2017neural,kang2018recommendation} 
or are available from additional sources~\cite{chen2016learning,nikolakopoulos2019recwalk}. 
More generally,
% In the latter case,
we assume that item features can be cast into a latent representation.
% (potentially using matrix factorization methods)
% using principal component analysis.
% \GW{is this correct? or does this apply to the first case?}
%Denoting the element-wise product by $\otimes$, 
Eq.~\ref{eq:pseudo-item} defines the 
pseudo-item $\vec{v_{ij}}$ as the element-wise geometric mean of $\vec{v_i}$ 
and $\vec{v_j}$. Compared to the arithmetic mean, the geometric mean 
is more appropriate for boosting similarities and dampening dissimilarities
(higher and lower values in the original vectors, respectively).  

The original LP algorithm requires an affinity matrix $W$ which encodes item-item similarities. 
In our problem, the labels we propagate are feedback points on $(v_i, v_j)$ pairs:
so each pseudo-item $v_{ij}$ represents a pair of items and the affinity matrix 
thus contains pair-pair similarities. This makes $W$ huge ($W \in \mathbb{R}^{|I|^2 \times |I|^2}$)
%near-futile w.r.t full materialization.
and prohibits full materialization.

Our approach rather is to materialize and populate merely a small subset of $W$
%Presenting the similarity function by $sim(.,.)$, 
% Therefore, we populate matrix $W$
by considering only
%$sim(v_{ij}, x)$ the 
the $k$ nearest neighbors of each pseudo-item $v_{ij}$ (a labeled feedback point).
A naive approach would require 
the generation of all 
possible pairs of items in which the nearest neighbors are computed
(with complexity $|I|^2)$).

To avoid this bottleneck, we compute an approximate 
$kNN$ set for each pseudo-item $v_{ij}$ using the following technique.
We find the $kNN$ set of $v_{ij}$ rather among the items in $I$,
denoted by the itemset $kNN^I_{ij}$ (the superscript $I$ denotes
that this is a set of items and not item-pairs or pseudo-items).
Instead of searching
in $|I| \times |I|$ for the $kNN$ of $v_{ij}$,
we search in $kNN^I_{ij} \times kNN^I_{ij}$.
This computation is made efficient using \textit{locality sensitive hashing (LSH)}
to deal with the large number of pairings.
% limiting the search space from $I \times I$ 
% to $C_{ij} \times C_{ij}$ where $|C_{ij}| << m$. The set $C_{ij}$ contains $c$ nearest neighbors 
% of $v_{ij}$
% among the items in $I$. 
% Pairing up the items in $C_{ij}$, we search for $kNN$s of $v_{ij}$ in $C_{ij} \times C_{ij}$ instead 
% instead of $I \times I$.
%
%This overall densification sequence reduces 
This way,
the search space for label propagation is reduced from $O(|I|^2)$ to $O(|I|)$.
%and highlights why \elixir is an efficient procedure.
%
% Throughout the described procedure, we always use locality sensitive hashing (LSH) for 
% identifying the nearest neighbors of an item. 
% \GW{this sounds like LSH is an additional accelerator - but isn't it already needed in the previous step when you talk about c nearest neighbors? how else would 
% you determine Cij ???}

To determine the $kNN$s of $v_{ij}$, an item-item
% distance or
similarity measure is required. 
Different recommenders use different measures for capturing
such
% item-item
similarities: cosine similarity~\cite{nikolakopoulos2019recwalk},
Euclidean %-based 
similarity~\cite{hsieh2017collaborative}, 
weighted inner products~\cite{kabbur2013fism, nikolakopoulos2019eigenrec, xin2019relational},
and angular similarity~\cite{sarwar2001item} 
are a few
% among the
common choices. % for determining item-item similarities. 
We treat the similarity function as a plug-in module $sim(.,.)$,
and instantiate it using cosine similarity when required.
% of given latent feature vectors
% in the rest of this 
%section
% paper.
Cosine similarity is emerging as a particularly convenient choice
when items (and often users, categories, etc.) are represented
in a shared latent space.
% \GW{should we emphasize latent representations here?}
Note that we treat items and pseudo-items uniformly and 
use the same function $sim(.,.)$ to compute their similarity. 
The output of this stage is a densified feedback matrix
$F^d_u$.

\subsection{Feedback incorporation}
%\textbf{Feedback Incorporation.} 

\textbf{Optimization problem.}
We incorporate matrix $F^d_u$ into the recommender by 
%adjusting items' latent 
%features. 
%For this, we 
imposing a soft constraint for learning a user-specific 
mapping function $g(., .)$ with 
 $\vec{w_u}$ as its parameter vector.
The goal is to learn preference vectors $\vec{w_u}$ for each user
that can be combined with existing item representations $\vec{v}$
to produce user-specific item representations and then
fed into the underlying recommender model.
To learn $\vec{w_u}$, we formulate the following objective function
where the signals from the densified feedback matrix are
incorporated as a soft constraint:
\begin{align}
	\min_{\vec{w_u}} & \frac{1}{m}\sum_{v_i, v_j} F^d_u(v_i, v_j) \cdot (sim(\vec{v_i}, \vec{v_j}) - 
	sim(g(\vec{v_i}, \vec{w_u}), g(\vec{v_j}, \vec{w_u}))) \nonumber \\
	& + \gamma ||\vec{w_u}||^2
	\label{eq:gen-feedback-inc}
\end{align}
%The above constraint can be simply added to the objective function as a regularizer and jointly 
%minimized through the optimization process in periodical model retraining. 
%Along with these constraints, 
where $m = |\{(v_i, v_j) | F^d_u(v_i, v_j) \neq 0\}|$. 
Eq.~\ref{eq:gen-feedback-inc} includes
%we select 
a mapping or transformation function 
$g$ (common choices would be vector translation and scaling), to be
computed by the optimization solver.
This serves to map original item representations and
re-arrange their positions in the latent space such that their new similarities 
reflect the user's feedback on item pairs. The underlying intuition is to decrease or
increase pairwise similarities whenever $F^d_u(v_i, v_j)=-1$  or $F^d_u(v_i, v_j)=1$, respectively.
The above objective achieves this two-fold (increasing and decreasing) effect in a unified manner.
Additional L2 regularization is used on $\vec{w_u}$ to encourage small magnitude,
avoiding drastic changes when it is applied inside \textit{g()}.

After learning an (near-) optimal $\vec{w_u}$,  each item vector 
$\vec{v_i}$ is updated to $g(\vec{v_i}, \vec{w_u})$; we use these 
user-specific item vectors 
to generate new recommendations. 
We posit that such user-specific item representations helps
the recommender
model to incorporate the \textit{more liked} and \textit{less disliked}
latent aspects of similarity for each user, and helps produce
improved recommendations.

An alternative choice of formulating Eq.~\ref{eq:gen-feedback-inc}
would be to have only the L2 regularization term as the objective
and model the signals from $F$ as hard constraints. 
With this alternative approach, the constraints would become inequality constraints
($F^d_u(v_i, v_j) \cdot (sim(\vec{v_i}, \vec{v_j}) - 
sim(g(\vec{v_i}, \vec{w_u}), g(\vec{v_j}, \vec{w_u}))) < 0$)
and would require that the KKT (Karush-Kuhn-Tucker) conditions be satisfied for
an optimal solution to exist. In practice, experimenting with 
the hard constraint formulation resulted in null solutions 
for most cases; hence our soft-constraint-based method. 

\textbf{Implementation.} The optimization in Eq.~\ref{eq:gen-feedback-inc}
for learning $\vec{w_u}$
is non-convex due to the presence of the cosine function.
Stochastic gradient descent (SGD) (available in libraries
like PyTorch) is used for computing near-optimal solutions.
% \GW{How does the learning of g work? gradient descent? need to say more here!}
% !TeX root = ../2021-www-fp-elixir.tex
\subsection{\elixir in recommenders with RWR}
\label{subsec:elixir-rwr}

\textbf{Generating recommendations.}
%PageRank 
%%has long been used as a 
%is a classical paradigm for ranking nodes in graphs
%%ranking mechanism in graphs~
%\cite{page1998, bahmani2010fast}. 
We incorporate our method into \recwalk~\cite{nikolakopoulos2019recwalk},
a state-of-the-art recommender model based on random walks with restart.
% (equivalently, Personalized PageRank).
% PageRank-based method for generating 
% recommendations.
The input to this model is a heterogeneous
graph (also referred to as a heterogeneous information network, HIN)
$G = (N, E, \theta)$ with a set of nodes $N$, a set of edges
$E \subseteq N \times N$ and a mapping $\theta$ from nodes and edges 
to their types, such that $\theta_N: N \mapsto T_N$ and $\theta_E : E \mapsto T_E$, 
where $|T_N| + |T_E| \ge 2$. Nodes are either of type $user$ or $item$, i.e., 
$N = U \cup I$. Edges capture user-item interactions, denoted by $A\in\{0, 1\}^{|N| \times |N|}$,
% ($n = |V|$)
and node-node similarities presented by 
$S \in \mathbb{R_+}^{|N| \times |N|}$.
So we have two types of nodes and two types of edges 
in this graph. 

In \recwalk, the recommendation score of item $v_i$ for user $u$ is computed 
as $PPR(u, v_i)$. $PPR$ stands for personalized PageRank~\cite{DBLP:journals/tkde/Haveliwala03},
%\GW{add reference, see comment}
%and $PPR(u, .)$ is 
%computed 
defined
as follows:
\begin{equation}
% 	PPR(u, .) = \alpha \overrightarrow{e_u} + (1-\alpha)PPR(u, .)(\beta A + (1-\beta)S)
		\vec{PPR(u, .)} = \alpha \cdot \vec{e_u} + (1-\alpha) \cdot \vec{PPR(u, .)} \cdot [\beta A + (1-\beta)S]
	\label{eq:recwalk}
\end{equation}
where $\alpha$ is the restart probability, $\vec{e_u}$ is the one-hot vector for user $u$ 
and $\beta$ is the probability that a walk traverses an interaction 
edge. According to Eq.~\ref{eq:recwalk}, a walk either visits one of its 
neighbors with probability $1-\alpha$ or jumps back to user node $u$. The neighbors are 
connected either through interaction or similarity edges. Matrix $S$ encodes similarities between 
nodes of the same type. Without loss of generality, we assume that a user is similar only to 
herself, i.e., $S(u_i, u_j)=1$ if and only if $i=j$. The item-item similarity, however, 
is defined by the $sim(.,.)$ function and hence $S(v_i, v_j) = sim(v_i, v_j)$. 
% As user-user similarities remain the same,
We simplify the notation 
and use $S$ to refer only to item-item similarities. Note that \recwalk normalizes matrix $S$ in a certain way to 
enforce stochasticity. We omit the details for the sake of brevity and refer users to~\cite{nikolakopoulos2019recwalk}
for more information.
The item $v$ in $|I|$ that has the highest $PPR(u, v)$ score, is produced
as the recommendation $rec$ for the user $u$.

\textbf{Generating explanations.}
Suppose item $rec$ is recommended to user $u$. Item-level explanations $\{exp\}$ in 
RWR-recommenders can be generated using the
%recent 
\prince algorithm~\cite{ghazimatin2020prince}. 
The resulting explanation item sets are minimal and counterfactual:
they ensure causality relation using the counterfactual setup
that $u$ would not have received $rec$ if she had not liked
the items in $\{exp\}$ in her history $H_u$.
However, minimality of explanations is
%of the explanations is 
not a concern in the current context.
Therefore, we take a more straightforward approach 
to approximate 
% the contribution scores of items in the user's action history.
\prince
by estimating
the contribution score of  item $v_j \in H_u$ to the recommended item $rec$:
\begin{equation}
	contribution(v_j, rec) = PPR(v_j, rec) \; \; \;(v_j \in H_u)
	\label{eq:rwr-contribution}
\end{equation}
where $PPR(v_j, rec)$ is the PageRank of node $rec$ personalized for node $v_j$. We use 
the top-$k$ items with highest contributions in $H_u$ as the explanation
set $\{exp\}$ for item $rec$. 

\textbf{Incorporating feedback.}
In \recwalk, item-item similarities are explicitly captured in matrix $S$, i.e., 
$S(v_i, v_j) = sim(v_i, v_j)$. 
%The construction of $S$ 
%can be approached in different ways. 
Given the items' latent representations (possibly computed by running
techniques like NMF or SVD from sparse explicit feature vectors),
% from an orthogonal  source,
we define $sim(v_i, v_j)$ as the cosine similarity between $v_i$ and $v_j$, and hence
$S(v_i, v_j)=cos(\vec{v_i}, \vec{v_j})$. 
As discussed earlier, to incorporate densified feedback $F^d_u$, we introduce a user-specific preference 
vector $\vec{w_u}$ to adjust $u$'s bias with respect to the latent aspects and
update the item representations by 
%simply 
adding $\vec{w_u}$. 
%Therefore, 
%the mapping function $g$ is defined as: 
The transformation function $g$ is chosen to be a vector translation,
shifting universal item representations onto user-specific ones:
\begin{equation}
	g(\vec{v_i}, \vec{w_u}) = \vec{v_i} + \vec{w_u}
	\label{eq:rwr-g}
\end{equation}

%\begin{figure} [t]
%	\begin{subfigure}{0.45\columnwidth}
%		\includegraphics[width=\columnwidth]{images/scale.png}
%		\caption{}
%		\label{fig:scale} 
%	\end{subfigure}
%	\begin{subfigure}{0.45\columnwidth}
%		\includegraphics[width=\columnwidth]{images/translation.png}
%		\caption{}
%		\label{fig:trans}
%	\end{subfigure}
%	\caption[Text]{Comparing (a) feature scaling and (b) feature translation (b) in 2D.}
%	\label{fig:g-compare}
%\end{figure}
%
%\GW{I find this argument not so clear, and the figure does not convey much to me. Consider dropping the figure and this paragraph!}\\
%We argue that translation is a better choice here than a scaling transformation,
%referring to a 2-D visualization of synthetic RWR data in Figure \ref{fig:g-compare}.
%With scaling the amount of expansion is limited, unlike translation.

The 
%primary idea behind 
intuition behind
the mapping described in Eq.~\ref{eq:rwr-g} is 
to highlight (suppress) the effect of liked (disliked) features 
through addition of positive (negative) bias values. 
% In the experimental section, we explore another method of translation based on 
% feature-scaling and
% provide insights on its limitations compared to the mapping defined in~\ref{eq:rwr-g}.  
% \GW{this point is not clear to me - translation of ... what exactly? item vectors into per-user item vectors?}
Plugging the definitions for $g$ and $sim$ and the densified matrix $F^d_u$ into the optimization objective (Eq.~\ref{eq:gen-feedback-inc}), we learn 
$\vec{w_u}$ as follows:
\begin{align}
	\min_{\vec{w_u}} & \frac{1}{m}\sum_{v_i, v_j} F^d_u(v_i, v_j) \cdot (cos(\vec{v_i}, \vec{v_j}) - 
	cos(\vec{v_i} + \vec{w_u}, \vec{v_j} + \vec{w_u})) %\nonumber \\ 
	%&+ 
	+\gamma ||\vec{w_u}||^2 
	\label{eq:rwr-feedback-inc}
\end{align}
Using $\vec{w_u}$, we build a user-specific similarity matrix $S_u$ defined as:
\begin{equation}
	S_u(v_i, v_j) = cos(\vec{v_i} + \vec{w_u}, \vec{v_j} + \vec{w_u})
	\label{eq:rwr-sim}
\end{equation}
Finally, we update the personalized PageRank recommendation scores accordingly, thereby completing the integration of pairwise feedback
into the recommender model:
\begin{equation}
	\vec{PPR(u, .)} = \alpha \cdot \vec{e_u} + (1-\alpha) \cdot \vec{PPR(u, .)} \cdot [\beta A + (1-\beta)S_u]
	\label{eq:recwalk-u}
\end{equation}

\section{User Study for Data Collection}
\label{sec:user-study}

%no ethics concerns were violated, and we had consent from users for 
%research participation, no personal data being made public, something like that.
%5. User study (2.5 pages)
%5.1 Data collection
%5.2 Evaluation results
%5.2.1 Key findings
%5.2.2 Analysis

\begin{table} [t] \small
	\centering
 	\resizebox{\columnwidth}{!}{
		\begin{tabular}{l c c c c}
			\toprule
			\textbf{Scenario}  &	\textbf{\#Item feedback}	&	\textbf{\#Pair feedback}	&	\textbf{Sessions}	&	\textbf{Hours} \\ \toprule
			Phase 1	& 	$50$			& 	$-$			&	$1$			&	$2$			\\
			Phase 2	& 	$30$			& 	$300$			&	$5$			&	$10$			\\ 
			Phase 3	& 	$72$		& 	$-$			&	$1$			&	$2$			\\ \midrule		
			Total	& 	$152$			& 	$300$			& $7$			& $14$			\\ \midrule
			Total (All users)	& 	$\simeq 4000$			& 	$7500$			& $175$			& $350$			\\ \bottomrule
	\end{tabular}}
	\caption{Annotations per user over stages of the study (spanning a total of $350$ person-hours) in each domain.}
	\label{tab:error}
	\vspace*{-0.3cm}
\end{table}

\elixir operates in a unique framework of user judgments on 
similarities between recommendation and explanation pairs.
It hinges on longitudinal observations
of the same users providing: i) original profiles,
ii) feedback on $rec$ items and iii) feedback on $\langle rec, exp \rangle$ pairs, 
as well as iv) item-level assessments on the final recommendations.
Thus, a study involving real users was imperative to
demonstrate the practical viability of our proposal.
%Since there is no publicly available test 
%collection with gold judgments on pairs of recommendations and explanations, 
%we designed a study with real users for data collection and subsequent
%evaluation.
To this end, we recruited $25$ volunteers, who were all
Masters' students of the Computer Science Department at
the authors' Institute, with payment per hour comparable to
that of master workers on a crowdsourcing platform
like Amazon Mechanical Turk.

\subsection{Statement on ethics}
\label{subsec:ethics}

Participants' privacy was fully respected:
all personally identifying information
concerning participants
was kept private during 
the course of the study, and deleted after its completion. 
All data was stored locally, with encryption, firewall protection and other measures of this sort.
During the course of the study, users had to provide ratings
on individual as well as pairs of movies and books. While this is not personally sensitive
per se, we recognize that the data reflects users' personal preferences.
All participants signed a consent document that they agree to this data being used for research purposes
and that it can be released with anonymized identifiers. 
%Users were never
%placed at risk and there were no deceptive practices involved. 
%GW: sounds overly defensive to me
%The proposal
%for the study was also submitted for approval by the authors' institutional
%ethics review board, and there were no objections.
The user study and the inclusion of results in this paper were approved
by the Ethics Review Board at the authors' Institute.

%Since physical gatherings are difficult,
The annotation sessions were conducted over online video conferencing,
so that participants' browser activity could be monitored.
To respect users' privacy, no video recordings were made.
% for these sessions.
A one-hour training session was conducted, where participants were 
made aware of the goals of the study and their exact tasks,
and were guided through examples.

\subsection{Setup}
\label{subsec:setup}

The user study was conducted in two domains: (i)  
movies (restricted to Hollywood, because of its popularity and the users' familiarity), and (ii) books. 
Over the course of six weeks (three weeks for each domain), 
each user annotated individual as well as pairs of movies and books 
for a total of $28$ hours.
%over a period of three weeks. 
The payment was $10$ Euros per hour, with the total cost amounting
to $25 \times 28 \times 10 = 7,000$ Euros.
For each domain, the annotations were collected in three phases that lasted three weeks.
%, roughly
%corresponding to the three weeks. % that the study spanned.
%We now describe the three stages in detail.
%It was also necessary that the study be completed in
%a reasonably short 
%stretch of time, to ensure that users retain familiarity with their
%interaction histories and to avoid
%a change in opinions or interests.

% users for total of $16$ hours over a period of $3$ weeks. The study was 
% conducted only in movie domain and had the following $3$ phases: 

\subsubsection{Phase 1: Building users' profiles}
\label{subsubsec:phase1}

It is essential to keep the assessment setup natural:
if users were asked to rate arbitrary items and pairs that they are unfamiliar
with, the judgments would be unreliable.
Thus, as the first step of the study for each domain, 
we asked users to provide 
us with $50$ movies and books each,
that they liked, to build a true history for each user, that would
create subsequent recommendations for her. Since movie or book titles can often be
ambiguous,
%(\textit{Murder on the Orient Express} has been made and re-made
%three times), 
users were asked to provide us with MovieLens (\url{https://movielens.org/home}) and Goodreads (\url{https://goodreads.com/}) URLs in individualized spreadsheets. 
% Similarly, for books, 
% we initiated users' profiles with $50$ books they liked. To remove duplicate items, 
% we asked users to share the Goodreads URLs (\url{https://https://www.goodreads.com/}) 
% of the books. Each book was then identified
% with the unique ID appearing in its URL. 
For each domain, we conducted this phase 
in a session spanning two hours which provided us
with $50 \times 25 = 1,250$ user actions % (i.e., likes).
(likes).

\subsubsection{Phase 2: Collecting feedback on items and pairs}
\label{subsubsec:phase2}

The obtained user profiles were plugged into
the RWR-based recommender model \recwalk~\cite{nikolakopoulos2019recwalk}, 
where
every liked item contributes an interaction edge to the network.
The union of all items rated by the $25$ users forms our universe
of items now, from where we generated the top-$30$ recommendations for
each user. Along with each recommendation, we generated
the top-$5$ explanation items $\{exp\}$ using
the approximated version of the \prince algorithm~\cite{ghazimatin2020prince} (Eq.~\ref{eq:rwr-contribution}).
To investigate the role of faithful explanations in pairwise feedback,
we also identified the five items $\{rand\}$ in the user's profile that are
the \textit{least similar}
to the recommendation item.
These serve as a proxy for pairing the recommendation with random items;
they are drawn from the user's profile to ensure familiarity.
The similarity is computed
as the cosine between the item vectors.

The users are now presented with three tasks:
(i) rate the generated recommendations (like/dislike);
(ii) rate the similarity of each $\langle rec, exp \rangle$ pair (like/dislike);
(iii) rate the similarity of each $\langle rec, rand \rangle$ pair (like/dislike).
The two kinds of feedback, item-level in (i), and
pair-level in (ii)+(iii),
have very different semantics, and users were appropriately briefed 
and guided. 
Item-level feedback is straightforward, where they comment whether they
liked or disliked an item.
Rating an item pair, though, needs a bit more reflection
on the possible similarities between the two items (two movies or two books), deciding on
the most important factor in case of multiple such aspects,
and providing the binary preference assessment. Participants entered their ratings 
in individualized spreadsheets we prepared for them. Each sheet contained several blocks 
where each block corresponded to one recommendation item followed by 
ten different explanation items for it 
(five $\{exp\}$ and five $\{rand\}$). To avoid any position bias, we randomly shuffled 
the explanation items in each block.  

While the feedback remains lightweight due to the user's potential
familiarity with the items,
%(for e.g. all \textit{rec} are somehow related to the \textit{exp}), 
we provided some help to cue their memory. For instance, 
we presented each movie with its title and its corresponding MovieLens URL, 
where the user could see
the movie's summary and key properties.
Moreover, MovieLens provides a rich set of tags on actors, directors,
genre, storyline and content; for item pairs we displayed
 the intersection set of top tags for the two movies.
%(this was done by meticulously scraping the Web pages with 
%Selenium-based browser automation).
%However, this was not meant to be exhaustive: users could still
%perform additional Web searches if they needed to brush up further facts,
%especially if the tag intersection turned out to be empty.
Users could nevertheless browse the MovieLens pages or other background sources
at their discretion. Book recommendations were also presented together with some auxiliary 
information including their descriptions, authors, top genres as listed on 
their Goodreads pages and their corresponding URLs. Similar to the movie domain, we 
facilitated users' judgments on pairs of books by listing their common properties such as genres and authors. 

Note that the
assessment of $\langle rec, exp \rangle$ or $\langle rec, rand \rangle$ is
decoupled from the fact whether the user likes $rec$, $exp$ or $rand$
individually.
To make this distinction clear, the users were walked through several
reference annotations during the training session.
For qualitative analysis, we also asked users to 
optionally articulate the dimension that was the basis of their
similarity feedback. We report on this in the experimental section.

At the end of this stage, each user provided us with
$30$ item-level ratings ($rec$),
and $30 \times 5 \times 2 = 300$ pair-level ratings
(five pairs for each of
$\langle rec, exp \rangle$ and $\langle rec, rand \rangle$). 
Therefore, for each domain, we had a total
of $750$ distinct item-level feedback points and $7,500$ distinct
pair-level feedback points, for a total of $25$ users. This phase required
ten hours from each user: to avoid task fatigue, this was spread over
five two-hour sessions.
% over the second week.

\subsubsection{Phase 3: Collecting feedback on final recommendations}
\label{subsubsec:phase3}

In the last phase of the longitudinal user study, the collected
feedback is incorporated
% using the \elixir
into the \elixir framework to produce improved recommendations for every user.
{\em Item-level feedback} is cast into additional interaction edges
for the original graph recommender; {\em pair-level feedback} is incorporated 
using the procedure described in 
% \elixir learning
Sec.~\ref{sec:framework}.
In addition, we experimented with {\em combined feedback} incorporating both
item-level and pair-level.
% is a seamless superimposition
%of the two. 
These are the three top-level configurations in our experimental evaluation.

For incorporating pair-level feedback, there are two possibilities of
using either $exp$ or $rand$, altogether resulting in five variations of
the recommender model. These models were each
made to produce $30$ recommendations,
leading to $180$ items to be rated by each user
(five pair-level and one item-level strategy, thus a total of 6 strategies $\times \; 30 = 180$).
However, there were overlaps in the $rec$ sets across configurations.
At the end, a total of $\simeq 1,800$ ratings were collected from $25$ users in each domain
($\simeq 72$ per user).
%instead of the worst case requirement of $180 \times 25 = 4,500$.
This phase took two hours per user, on average, and was 
completed in a single session at the end of the third week.

\section{Evaluation}
\label{sec:results}

\subsection{Configurations}
\label{subsec:exp-setup}

% We are now in a position to evaluate \elixir and its compare the relative performance 
%of alternative mechanisms of incorporating feedback.
We evaluate \elixir for different configurations, including the baseline of
exploiting solely item-level feedback on $rec$ items. But before we can go to the results, we need to explain the basic setup of the experimental framework.
Latent vectors necessary to initialize item representations are learnt
by running non-negative matrix factorization (NMF) on the sparse matrix
of movie-tag memberships from MovieLens and 
book-genre memberships from Goodreads 
using the Nimfa Python library
(\url{http://ai.stanford.edu/~marinka/nimfa/}, default settings).
The number of latent dimensions $d$ is chosen to be $20$ which 
was guided by observations on the reduction of sparsity from the original matrix.
% In Sec.~\ref{sec:discussion}, we discuss how resource constraints hinder a through parameter grid search. 
We use the SciPy library
(\url{https://bit.ly/35lVV10})
for subsequent label propagation with the cosine kernel.
The number $k$ for
LP was chosen to be $10$, which means that for each pseudo-item, we find
the $10$ nearest items, and hence $ {10 \choose 2} = 45$ pseudo-items. We tried two other values, 
$k=5$ and $20$, and observed similar results.  
For LSH, we used NearPy
(\url{https://pixelogik.github.io/NearPy/})
with random binary projection as its 
hash function. 
% The number of random vectors is set to $3$ resulting in $8$ separate buckets. 
% LSH projects each item vector $v_i$ into one of these buckets where its approximate
LSH assigns each item vector $v_i$ to a bucket where its approximate 
nearest neighbors lie. While a large number of buckets decreases the probability 
of neighbors to be assigned to the same bucket, a small number reduces the efficiency of 
$k$NN queries. Considering the choice of $k$ in LP ($k=10$), 
we chose the number of buckets to be $8$ (corresponding to $3$ random binary projection vectors). 
With this number of buckets, we reduce the failure rate of LSH to $15\%$, i.e., for only $15\%$ of the $k$NN queries 
with $k=10$, LSH returns 
less than $10$ neighbors. 
We use our own implementations of \recwalk and \prince for
generating recommendations and explanations, respectively.

The following five feedback configurations are compared: 
\squishlist
\item \textit{Item-level feedback.}  This baseline model only absorbs users' binary preferences on individual items $rec$. 
Such item-level feedback adds interaction edges to the RWR graph.
\item \textit{Pair-level feedback with explanations.} The model captures only the 
judgments on pairs of $\langle rec, exp \rangle$ items. Such pair-level signals update the similarity matrix used in RWR.
\item \textit{Pair-level feedback with random items.} This is similar to the previous 
configuration, except that the explanation items here are replaced by the least relevant
items from the user's history ($\{rand\}$). % thus are not necessarily faithful. 
\item \textit{Item + pair-level feedback with explanations.} The model exploits both 
individual and pairwise feedback.
\item \textit{Item + pair-level feedback with random items.} This is similar to the previous configuration 
except that the explanation items $\{exp\}$ are replaced by $\{rand\}$.
\squishend

\subsection{Metrics}
\label{subsec:metrics}

We evaluate the quality of recommendations generated
after feedback incorporation using three metrics: i) Precision at the top-$k$ ranks (P@$k$),
ii) Mean Average Precision at the top-$k$ ranks (MAP@$k$), and, iii) normalized discounted cumulated gain at the top-$k$ ranks (nDCG@$k$,
computed with binary non-graded relevance assessments).
While P@$k$ is a set-based metric considering the top-$k$ items
analogous to a \textit{slate of recommendations}, the latter two 
are sensitive to the ranks of relevant items in the lists.
All metrics are computed at three rank cut-off values $k$: $3$, $5$, and $10$.
The relatively low values of cut-off ranks are chosen to show the effectiveness
of \elixir in introducing highly selective items into
the top recommendations for individual users.

\subsection{Initialization}
\label{subsec:init}

To fairly compare different configurations, we train \recwalk
using the same set of parameters. The restart probability $\alpha$ 
is set to $0.15$ as shown effective in prior works. To highlight 
the effect of similarity edges, we choose 
$\beta=0.1$: a lower $\beta$ indicates a lower likelihood
of sampling an interaction edge for the random walker, and
a walker thus traverses similarity edges in $G$ with probability $0.9$. 
Using smaller values for $\beta$ is also suggested in the original 
model of \recwalk~\cite{nikolakopoulos2019recwalk}. 
The interaction graph $G$ built for movies  
had $25$ users, 
$621$ movies, $1.3k$ interaction edges, and $11k$
similarity edges. For books, $G$ 
was larger and denser, with $868$ books, 
$1.3k$ interaction edges and $41k$ 
similarity edges. 
To compute PageRank scores, we use the power-iteration method with
a maximum of $500$ iterations. 

% We fix the number of latent 
% features to $20$ throughout our experiments. 
% Items' representations are learned by 
% applying NMF on concrete feature vectors. 
To construct the similarity matrix $S$ for \recwalk, 
we employ LSH again with a similar configuration as discussed for densification. 
To avoid too many edges in the graph, we only connect items 
with large similarity. For this, we use threshold $0.7$, i.e., $S(v_i, v_j) = 0$ 
if $cos(v_i, v_j) < 0.7$. Matrix $A$ is built from item-level user feedback:
we define $A(u_i, v_j) = 1$ if $u_i$ likes
item $v_j$, and zero otherwise.
Regularization parameter
$\gamma$ in Eq.~\ref{eq:rwr-feedback-inc},
and the learning rate $lr$
in SGD
for finding an optimal $\vec{w_u}$
were tuned on a development set containing $20\%$ of each user's ratings. 
We tested $10$ different values for $\gamma$ ($1, 2, \ldots 10$) and $3$ values for the learning rate 
($0.001, 0.01, 0.1$) and chose the values with best performance on the development set: $\gamma=3$ and $lr=0.01$.
%best value for $\gamma$ and the learning rate [1-10]to be $3$
%and $0.01$ respectively.

\begin{table*} [t]
	\centering
	\begin{tabular}{l | ccc | ccc | ccc} 
		\toprule
		\textbf{Setup [Movies]}                        & \textbf{P@3}      & \textbf{P@5}      & \textbf{P@10}     & \textbf{MAP@3}    & \textbf{MAP@5}    & \textbf{MAP@10}   & \textbf{nDCG@3}   & \textbf{nDCG@5}   & \textbf{nDCG@10}   \\ 
		\toprule
		Item-level                   & $0.253$  & $0.368$  & $0.484$  & $0.323$  & $0.380$   & $0.448$  & $0.244$  & $0.327$  & $0.422$   \\ \midrule
		Pair-level (Exp-1)         & $0.506$* & $\boldsymbol{0.592}$* & $0.580$* & $0.566$* & $0.599$* & $0.625$* & $0.496$* & $\boldsymbol{0.557}$* & $0.565$*  \\
		Pair-level (Exp-3)         & $0.506$*	& $0.568$*	& $\boldsymbol{0.596}$*	& $\boldsymbol{0.630}$*	& $\boldsymbol{0.624}$*	& $\boldsymbol{0.634}$*	& $\boldsymbol{0.504}$*	& $0.547$*	& $\boldsymbol{0.575}$*  \\
		Pair-level (Exp-5)         & $0.480$* & $0.512$* & $0.568$*  & $0.567$* & $0.579$* & $0.591$* & $0.463$* & $0.490$* & $0.536$*  \\
		Pair-level (Rand-5)      & $0.453$* & $0.504$*  & $0.560$*  & $0.536$* & $0.537$* & $0.596$* & $0.451$* & $0.486$* & $0.532$*  \\ \midrule
		Item+Pair-level (Exp-5)    & $\boldsymbol{0.533}$* & $0.544$* & $\boldsymbol{0.596}$* & $0.563$* & $0.571$* & $0.603$* & $0.500$* & $0.517$*  & $0.562$*  \\
		Item+Pair-level (Rand-5) & $0.453$* & $0.488$* & $0.572$* & $0.486$* & $0.520$* & $0.578$* & $0.426$* & $0.458$* & $0.526$*  \\
		%\midrule
		%\bottomrule
		\toprule
		\textbf{Setup [Books]}                        & \textbf{P@3}      & \textbf{P@5}      & \textbf{P@10}     & \textbf{MAP@3}    & \textbf{MAP@5}    & \textbf{MAP@10}   & \textbf{nDCG@3}   & \textbf{nDCG@5}   & \textbf{nDCG@10}   \\ 
		\toprule
		Item-level                   & $0.253$	& $0.336$	& $0.436$	& $0.309$	& $0.343$	& $0.417$	& $0.233$	& $0.296$	& $0.379$   \\ \midrule
		Pair-level (Exp-1)         & $0.506$*	& $0.528$*	& $0.54$*	& $0.603$*	& $0.620$*	& $0.654$*	& $0.493$*	& $0.512$*	& $0.527$*  \\
		Pair-level (Exp-3)         & $0.506$*	& $0.536$*	& $0.58$*	& $0.573$*	& $0.596$*	& $0.612$*	& $0.484$*	& $0.511$*	& $0.551$*  \\
		Pair-level (Exp-5)         & $\boldsymbol{0.586}$*	& $\boldsymbol{0.600}$*	& $\boldsymbol{0.656}$*	& $\boldsymbol{0.726}$*	& $\boldsymbol{0.701}$*	& $\boldsymbol{0.692}$*	& $\boldsymbol{0.602}$*	& $\boldsymbol{0.607}$*	& $\boldsymbol{0.644}$*  \\
		Pair-level (Rand-5)      & $0.480$*	& $0.504$*	& $0.504$	& $0.593$*	& $0.604$*	& $0.588$*	& $0.468$*	& $0.488$*	& $0.494$*  \\ \midrule
		Item+Pair-level (Exp-5)    & $0.493$*	& $0.456$*	& $0.560$*	& $0.543$*	& $0.566$*	& $0.566$*	& $0.482$*	& $0.458$*	& $0.529$*  \\
		Item+Pair-level (Rand-5) & $0.386$*	& $0.416$*	& $0.516$*	& $0.469$*	& $0.509$*	& $0.536$*	& $0.376$*	& $0.399$*	& $0.475$*  \\
		\bottomrule
	\end{tabular}
	\caption{Comparison of different modes of feedback incorporation with results from the user study. 
		The best value in every column (for each domain) is marked in \textbf{bold}. * denotes statistical significance of all methods over item-level feedback,
		with $p$-value $\leq 0.05$ using the Wilcoxon signed rank test~\cite{wilcoxon1992individual}.}
	\label{tab:main}
		\vspace*{-0.5cm}
\end{table*}

\subsection{Key findings}
\label{subsec:key}

Our key findings are presented in Table ~\ref{tab:main}, where different feedback absorption strategies are evaluted
over data from the user study. We make the following salient observations:
\squishlist
	\item \textbf{Pair-level feedback improves recommendations.} The most 
%convincing insight 
salient observation
from the results in both domains is that 
including pairwise
	feedback (all table rows except the first one 
	for each domain) 
	on the similarity of item pairs 
%can result in across-the-board improvement 
results in substantial improvements
	over solely considering item-level feedback. This confirms
	our hypothesis that procuring feedback on pairs provides 
%distinctive signals from each user on their likes and dislikes
%	(like filling in missing pieces of user profiles), 
refined and highly beneficial user signals
that cannot be captured through aggregated item-level feedback.

\item \textbf{Pair-level feedback is more discriminative than item-level.} Next, we compare 
the effectiveness of pair-level feedback vis-\'{a}-vis item-level. % for improving recommendations. 
To have a fair comparison with respect to volume of feedback, 
we introduce a new setup, Pair-level (Exp-1), where 
the volume of pair-level feedback is similar to that of the Item-level setup.
For this, we only consider users' feedback on the most relevant explanation item. This results in  
incorporation of $30$ pairs with distinct recommendation items, that is compared to item-level feedback on these $30$ items. We observe that in both 
domains, Pair-level (Exp-1) significantly outperforms the
Item-level setup. This demonstrates the 
efficacy
% effectiveness
of pair-level feedback in capturing users' fine-grained interests. For completeness, we also provide results when we use top-3 and top-5 explanations (Exp-3 and Exp-5, respectively). 

%\item \textbf{Item+Pair-level is the best choice.} Our study reinforces the traditional utility of learning from item-level feedback, as we observe
%that the best configuration across all metrics is the one where item-level feedback is combined with pair-level. 
%Thus, in practice, a reasonable
%	way of introducing such feedback could be to first judiciously ask for pair-level feedback, and then iteratively elicit more similarity assessments from users who find this mechanism comfortable. 
	\item \textbf{Using explanations for pair-level feedback is essential.} We set out with the goal of making explanations actionable towards model improvement. This is validated by the observation that item+pair-level for explanations are consistently and substantially better than item+pair-level for random item instead of top-ranked explanation items. 
%If we consider only pair-level feedback (that is somewhat less reliable than the joint configurations), we see that explanations can result in improvements for higher ranks (P@5, MAP@3, MAP@5). Nevertheless, it is heartening to see that the model can learn discriminative signals on user profiles even recommendations are paired with random items from the user's history.
%%%GW: this is a second-order point that diminishes the striking point about top vs. bottom/rand exp items - hence commented out
% \item \textbf{Item+Pair-level feedback increases the homogeneity of the recommendations the most.} Next, we explore the effect of incorporating different types of feedback on homogeneity of the top recommendations. For this, we first generate the top-$10$ recommendations for three setups: (1) Item-level, (2) Pair-level (Exp-$5$), and (3) Item+Pair-level (Exp-$5$). To measure homogeneity of the top recommendations, we compute the entropy of their tag distribution, where the smaller values are interpreted as larger homogeneity. The entropy values for setups $1$-$3$ in the movie domain are $6.409$, $6.371$ and $6.291$, respectively. Note that in the movie domain with the increase in homogeneity the recommender performs better. The entropy values in the book domain are $4.401$, $4.553$ and $4.345$, respectively for setups $1$-$3$. Unlike for movies, we observe that less homogeneity (corresponding to setup $2$) yields the best results which might be due to willingness of some participants to read more diverse books.
\squishend

\subsection{Analysis}
\label{subsec:analysis}

Our longitudinal study on collecting pairwise judgments opens up the possibility of 
gaining insights on several issues on user behavior and feedback:
%\squishlist
%	\item How different are positive and negative feedback by volume?
%	\item Are users who are more likely to provide negative item-level ratings also biased towards more dislikes on item pairs?
%	\item What are the most common aspects influencing similarity feedback on item pairs?	
%	\item How does performance improvement correlate with the diversity of the original profile?	
%	\item How do the volumes of positive and negative feedback correlate with performance improvement?
%\squishend

\begin{enumerate}[label=Q\arabic*.]
\item How different are positive and negative feedback by volume?
\item Are users who are more likely to provide negative item-level ratings also biased towards more dislikes on item pairs?
\item What are the most common aspects influencing similarity feedback on item pairs?
\item How does performance improvement correlate with the diversity of the original profile?
\item How do the volumes of positive and negative feedback correlate with performance improvement?
%\item How does pair-level feedback affect diversity of the recommendations?
\end{enumerate}

\begin{figure*}[t]
	\centering
	\begin{subfigure}{\textwidth}
		\includegraphics[width=\textwidth]{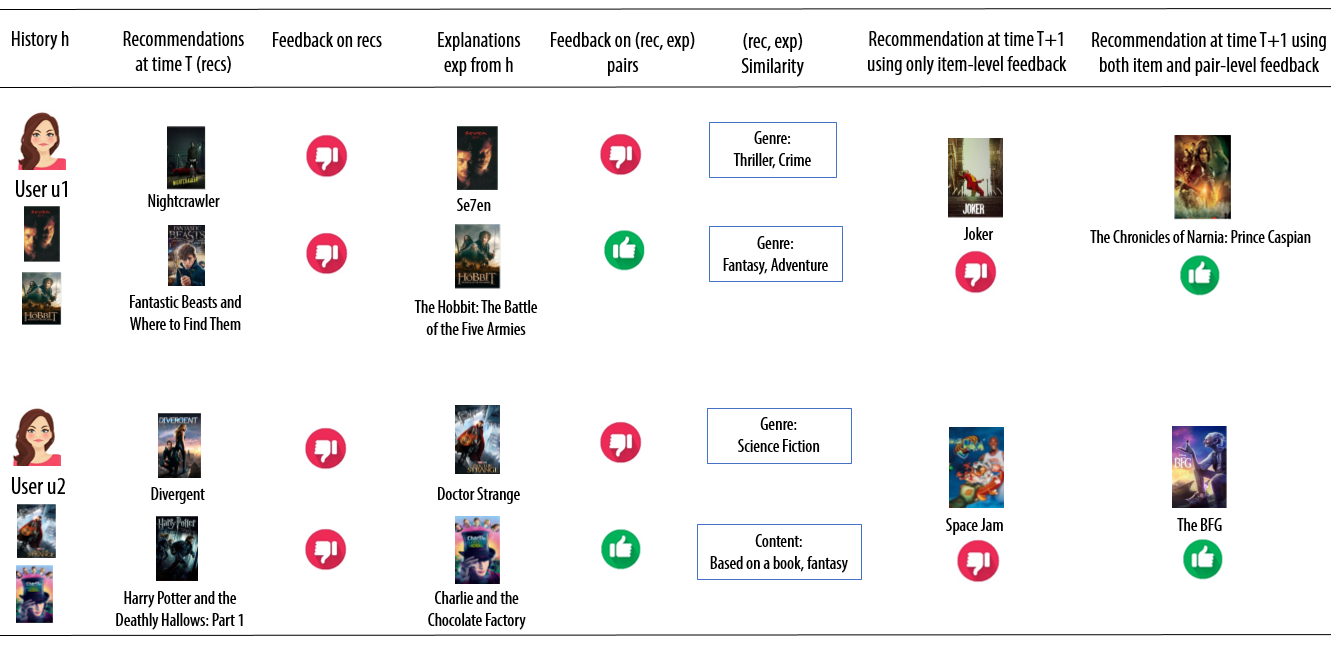}
		\caption{Movies}
		\label{fig:anecdote-movies} 
	\end{subfigure}	
	\begin{subfigure}{\textwidth}
		\includegraphics[width=\textwidth]{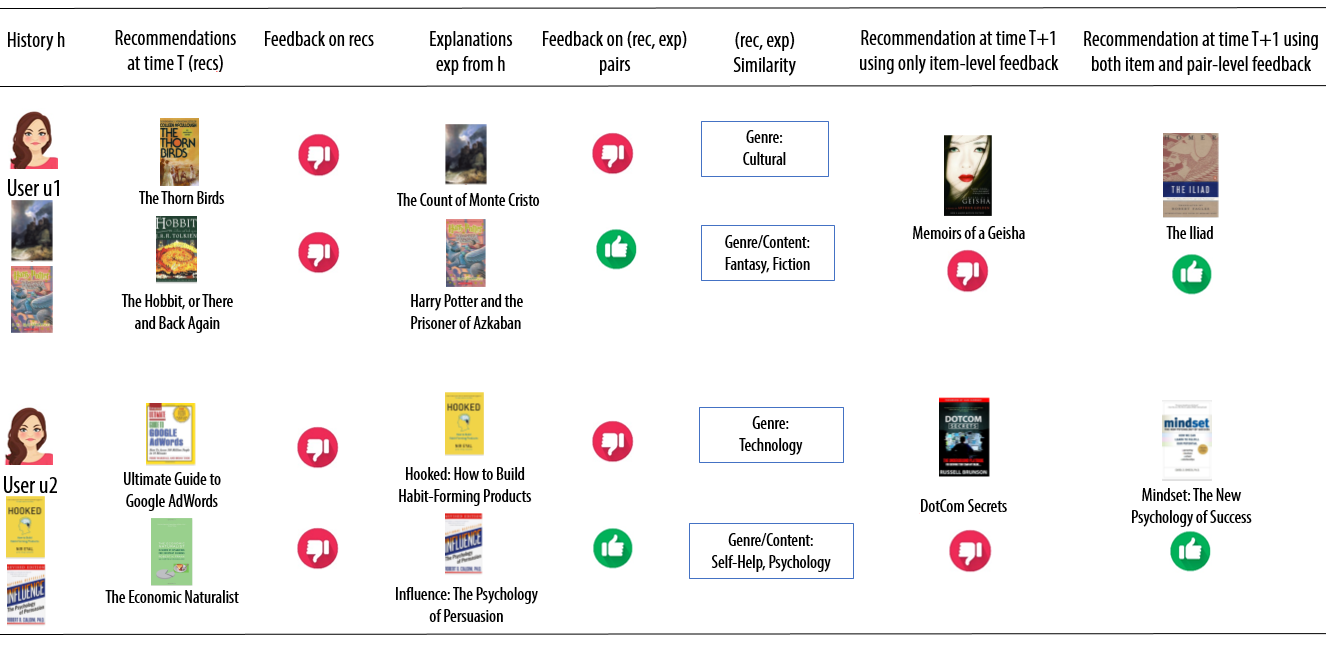}
		\caption{Books}
		\label{fig:anecdote-books}
	\end{subfigure}
	\caption[Text]{Anecdotal examples showing the effectiveness of \elixir in introducing relevant recommendations to the top-10.}
	\label{fig:anecdotes}
\end{figure*}

We address these questions in the subsequent analysis. At the end, Fig.~\ref{fig:anecdotes}
shows four 
representative examples from the user study.

\begin{figure}[t]
  \centering
  \begin{subfigure}{.495\columnwidth}
    \centering
    %\captionsetup{width=.9\columnwidth}
    \includegraphics[width=\linewidth]{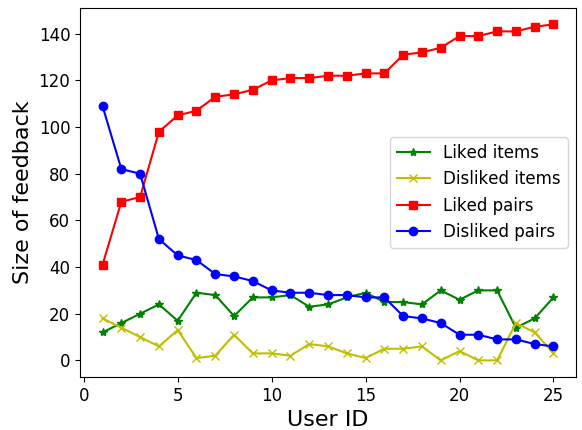}
    \caption{Movies}
    \label{fig:feedback-dist-movielens}
  \end{subfigure}%
  \hfill
  \begin{subfigure}{.495\columnwidth}
    \centering
    %\captionsetup{width=.8\columnwidth}
    \includegraphics[width=\linewidth]{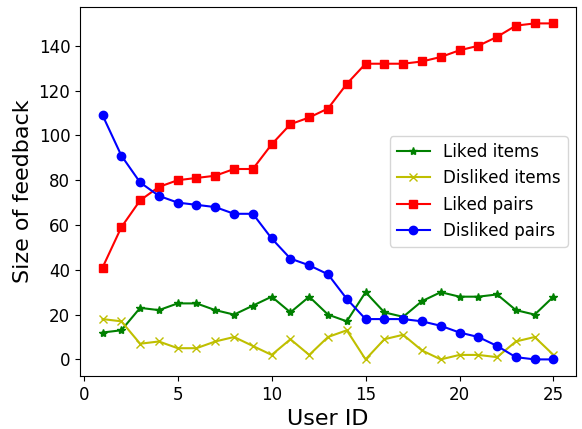}
    \caption{Books}
    \label{fig:feedback-dist-goodreads}
  \end{subfigure}%
  \vspace{-1\baselineskip}
  \caption{Per-user volume of feedback by type.}
  \label{fig:feedback-dist}
\end{figure}

To see size distribution of different feedback types (Q1), 
we plot the number of item-level and pair-level feedback points
provided by each user
in Fig.~\ref{fig:feedback-dist}. % An inherent disparity in the number of such pairs
%Under a null hypothesis, we would expect that all the four line plots remain parallel
%to each other by shape: however, that is not the case. 
Users vary in their proportions
of positive and negative feedback. 
%However, 
%as shown in Fig.~\ref{fig:feedback-corr}, 
%there is a strong correlation between 
%the number of positive (likes) feedback at item-level and the same number at pair-level 
%The nagative correlation between the positive and the negative feedback (at any level) is explained 
%by the fact that the feedback is binary and the total number of feedback is constant. 
%and their behavior does not exactly carry over 
%from items to pairs. 
Overall, users enter much more positive feedback than
negative. Reasonable numbers for all four types of judgments (item-/pair-level $\times$ like/dislike) show that, overall, users
are willing to provide the effort towards improving their recommendations.
Monitoring feedback assessments over online sessions
showed that pairwise feedback is indeed lightweight effort (measured by time taken for completion)
and does not impose severe cognitive load on users.

In particular, we find that users who may be
biased towards more negative item-level feedback often % tend to provide similar signals at a pair-level:
% they
provide substantial volumes of negative feedback on pairs. 
The corresponding Pearson correlations in 
movie and book domains are $0.53$ 
and $0.51$, respectively. 
Therefore, we can conclude that user behavior carries over from items to pairs (Q2).
This leads to an extremely crucial insight: negative signals on items cannot be harnessed in 
graph recommenders, and are simply discarded as bad recommendations, never to be used again.
Yet eliciting negative feedback on certain pairs of bad recommendations
and good explanation items,
can lead to substantial benefit for the recommender system: similar bad recommendations become 
less likely to be recommended to the user.
%This is the key idea
%of \elixir.

%\begin{figure}[t]
%	\centering
%	\includegraphics[width=\columnwidth]{images/like_dislike_dist.png}
%	\caption{Per-user volume of feedback by type.}
%	\label{fig:feedback-dist}
%		\vspace*{-0.5cm}
%\end{figure}
%\begin{figure}[t]
%	\centering
%	\includegraphics[width=\columnwidth]{images/heatmap.png}
%	\caption{Pearson correlation of different types of feedback.}
%	\label{fig:feedback-corr}
%		\vspace*{-0.5cm}
%\end{figure}

\begin{figure}[t]
    \centering
      \begin{subfigure}{.5\columnwidth}
        \centering
        \captionsetup{width=.9\columnwidth}
        \includegraphics[width=\linewidth]{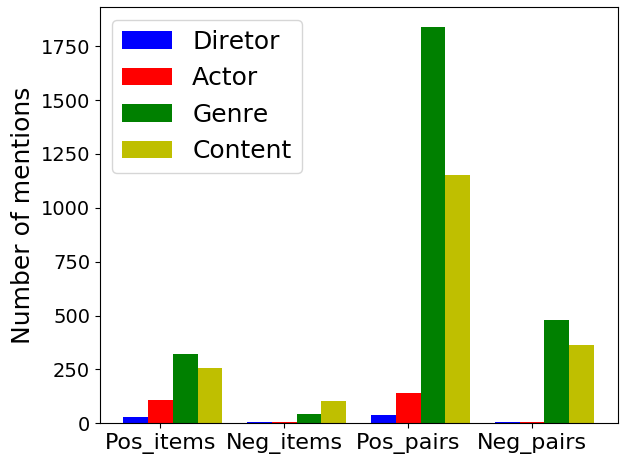}
        \caption{Movies}
        \label{fig:reason-dist-movielens}
      \end{subfigure}%
      \hfill
      \begin{subfigure}{.5\columnwidth}
        \centering
        %\captionsetup{width=.8\columnwidth}
        \includegraphics[width=\linewidth]{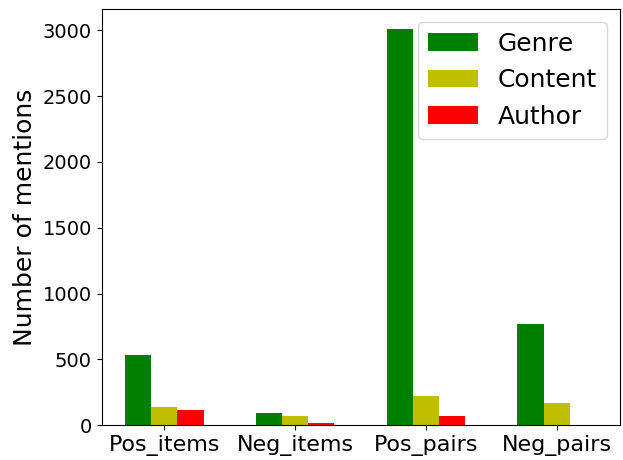}
        \caption{Books}
        \label{fig:reason-dist-goodreads}
      \end{subfigure}%
      \vspace{-1\baselineskip}
      \caption{Key influencers behind feedback assessments.}
	\label{fig:reason-dist}
	\vspace*{-0.5cm}
\end{figure}

Next, we show factors influencing pairwise similarity assessment in Fig.~\ref{fig:reason-dist} (Q3).
To compare and contrast, we asked users to mention their reasons for both 
item- and pair-level feedback. Qualitative analysis reveals that in the movie domain, genres
play the biggest role in feedback, followed by content, actor, and then director. 
We observe that genre and content (the latter includes storylines like plot twists,
alien movies, medieval movies, etc.) are much more likely to influence
user preferences than presence of specific actors or directors.
This underlines the necessity of latent representation of item properties, as storylines are
hard to capture in explicit feature models.
Similar trends are observed for books, % domain, 
i.e., genre is the most frequently mentioned, followed by content 
and author. 
The interesting observation is that users are systematic in their behavior:
in both domains, histograms have the same relative distribution 
for item- and pair-feedback.

\begin{figure}[t]
    \centering
      \begin{subfigure}{.5\columnwidth}
        \centering
        \captionsetup{width=.9\columnwidth}
        \includegraphics[width=\linewidth]{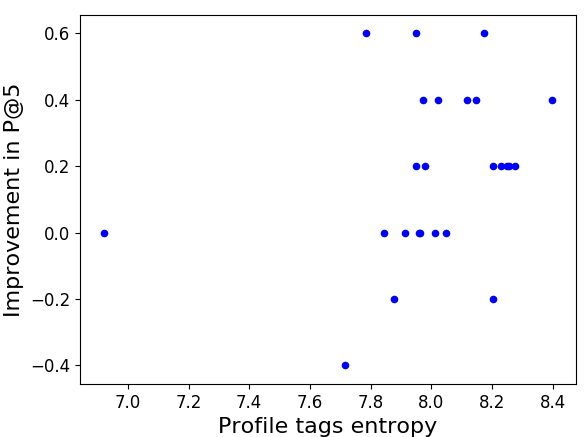}
        \caption{Movies}
        \label{fig:imp-entropy-movielens}
      \end{subfigure}%
      \hfill
      \begin{subfigure}{.5\columnwidth}
        \centering
        %\captionsetup{width=.8\columnwidth}
        \includegraphics[width=\linewidth]{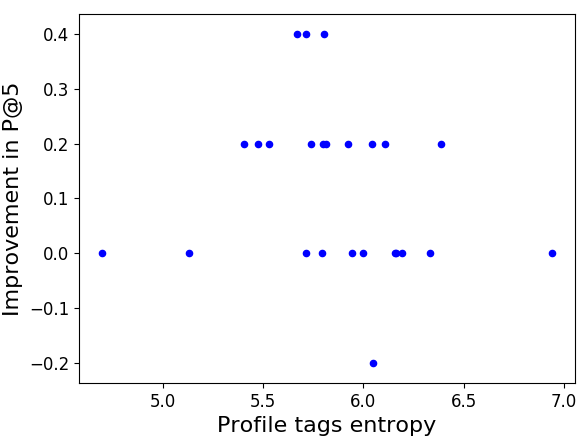}
        \caption{Books}
        \label{fig:imp-entropy-goodreads}
      \end{subfigure}%
      \vspace{-1\baselineskip}
      \caption{Connecting gains via \elixir with profile diversity.}
	\label{fig:imp-entropy}
	\vspace*{-0.5cm}
\end{figure}

We also investigate whether all users are equally likely to benefit
from \elixir. Since profile sizes are kept constant to control for
other factors, we try to see if 
performance improvements from item+pair-level
feedback is connected to the \textit{diversity} of their original profiles (Q4).
To quantify diversity, we measure the entropy of the distribution 
of tags associated with the $50$ items that were used to initialize
profiles of the users (higher entropy is higher diversity).
Plots for movies and books are shown in Fig.~\ref{fig:imp-entropy}. Our observation for the movie domain is that
\elixir helps users with relatively high interest diversity 
(right half) slightly more (top right) than the users with
more particular interests 
(languishing towards the bottom left corner). The corresponding 
Pearson correlation is $0.29$ which indicates 
moderate positive correlation 
between profile diversity and improvement level. For books, % domain, 
however, the Pearson correlation is $-0.17$ implying a 
small negative correlation between diversity of profiles and 
effectiveness of \elixir. 

\begin{figure}[t]
	\centering
	\begin{subfigure}{0.495\columnwidth}
		\centering
		\includegraphics[width=\columnwidth]{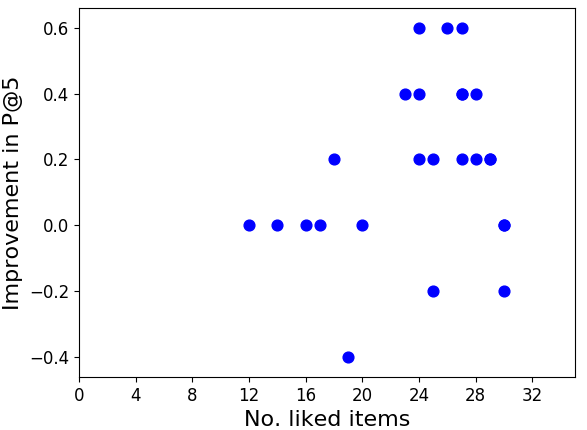}
		\caption{}
%		{{\small Network 1}}    
		\label{fig:imp-item-pos}
	\end{subfigure}
	\hfill
	\begin{subfigure}{0.495\columnwidth}  
		\centering 
		\includegraphics[width=\columnwidth]{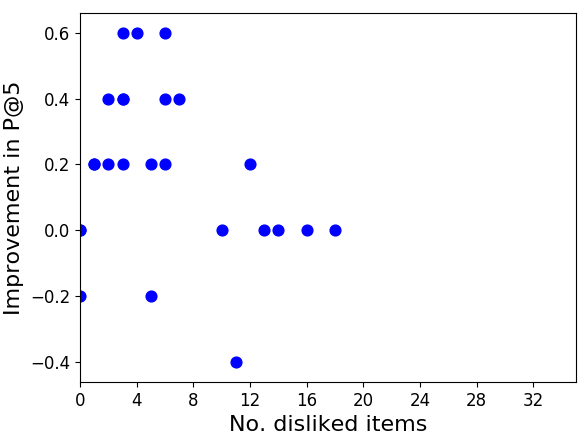}
		\caption{}
		%{{\small Network 2}}    
		\label{fig:imp-item-neg}
	\end{subfigure}
	%\vskip\baselineskip
	\vspace{-0.2\baselineskip}
	\begin{subfigure}{0.495\columnwidth}   
		\centering 
		\includegraphics[width=\columnwidth]{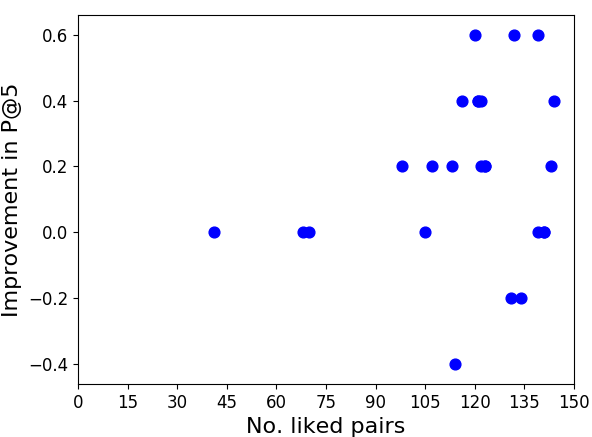}
		\caption{}
		%{{\small Network 3}}    
		\label{fig:imp-pair-pos}
	\end{subfigure}
	\hfill
	\begin{subfigure}{0.495\columnwidth}   
		\centering 
		\includegraphics[width=\columnwidth]{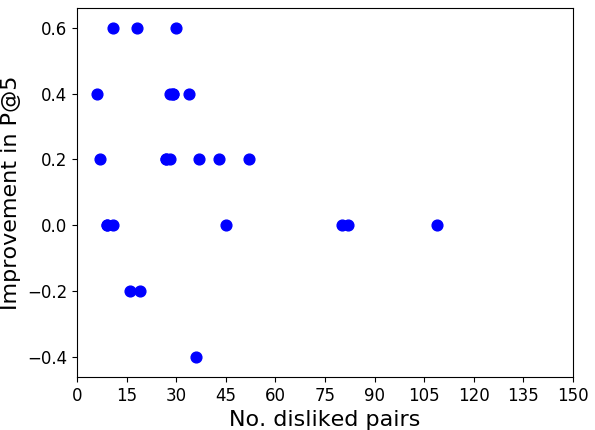}
		\caption{}
		%{{\small Network 4}}    
		\label{fig:imp-pair-neg}
	\end{subfigure}
		\vspace*{-0.3cm}
	\caption{P@5-improvement w.r.t. feedback size (Movies).}
	%{\small The average and standard deviation of critical parameters: Region R4} 
	\label{fig:imp-size}
		\vspace*{-0.3cm}
\end{figure}

\begin{figure}[t]
	\centering
	\begin{subfigure}{0.495\columnwidth}
		\centering
		\includegraphics[width=\columnwidth]{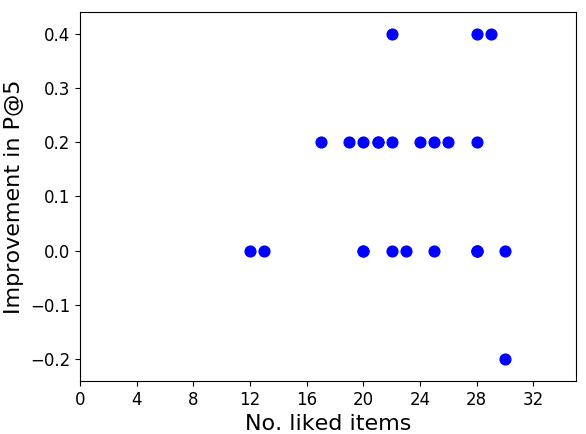}
		\caption{}
%		{{\small Network 1}}    
		\label{fig:imp-item-pos-goodreads}
	\end{subfigure}
	\hfill
	\begin{subfigure}{0.495\columnwidth}  
		\centering 
		\includegraphics[width=\columnwidth]{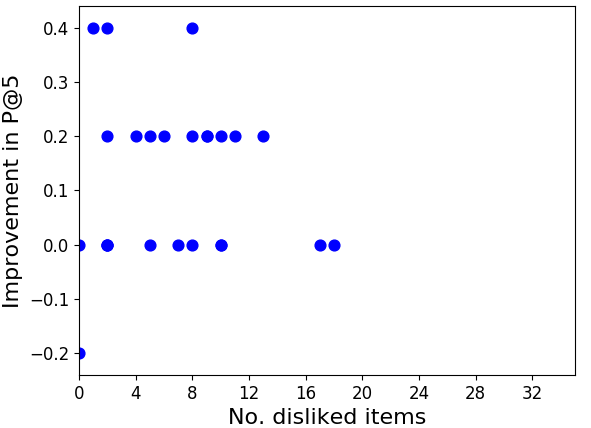}
		\caption{}
		%{{\small Network 2}}    
		\label{fig:imp-item-neg-goodreads}
	\end{subfigure}
	%\vskip\baselineskip
	\vspace{-0.2\baselineskip}
	\begin{subfigure}{0.495\columnwidth}   
		\centering 
		\includegraphics[width=\columnwidth]{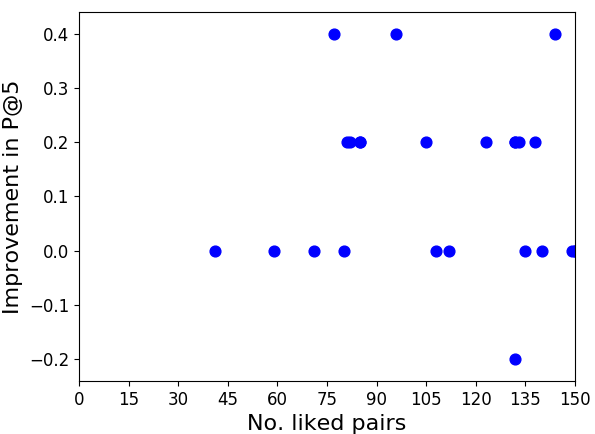}
		\caption{}
		%{{\small Network 3}}    
		\label{fig:imp-pair-pos-goodreads}
	\end{subfigure}
	\hfill
	\begin{subfigure}{0.495\columnwidth}   
		\centering 
		\includegraphics[width=\columnwidth]{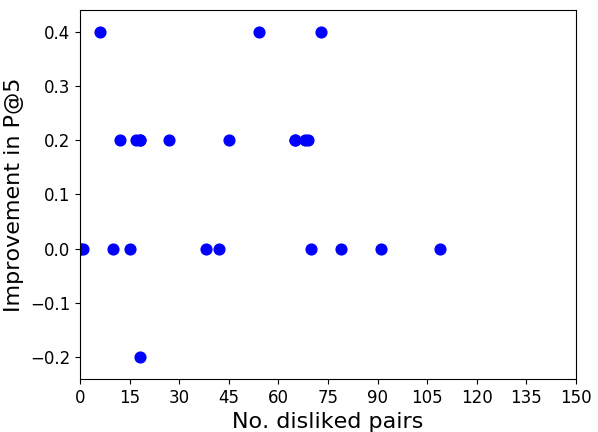}
		\caption{}
		%{{\small Network 4}}    
		\label{fig:imp-pair-neg-goodreads}
	\end{subfigure}
		\vspace*{-0.3cm}
	\caption{P@5-improvement w.r.t. feedback size (Books).}
	%{\small The average and standard deviation of critical parameters: Region R4} 
	\label{fig:imp-size-goodreads}
		\vspace*{-0.3cm}
\end{figure}

Next, we investigate if different volumes of feedback on the four
possibilities (item-level like/dislike; pair-level like/dislike)
lead to notably differing performance improvements (Q5). 
We show the effect of feedback size on improvement levels 
in Fig.~\ref{fig:imp-size} 
and~\ref{fig:imp-size-goodreads}, where 
the two plots in the top row of each figure 
correspond to item-level feedback,
and the bottom to pair-level. The scales and limits of $x$-axes within rows (and all $y$-axes)
are kept the same for easy comparison. 
Here we note
that dots along the same row (level) of precision correspond to 
the same users.
The notable observation from
% Fig.~\ref{fig:imp-size}
these figures is that % , for movies,
users who provide more 
positive
feedback are likely to see higher improvements. This correlation 
is particularly more pronounced for positive item-level 
feedback.
% (Pearson correlation$=0.30$).
While the benefit
of sending more
positive signals as more actionable is understandable, a certain
part of the blame may lie on the graph recommender itself,
where ``negative edges'' cannot be included easily: presence 
or absence of edges is the standard model. This suggests further research 
to explore \elixir with other families of recommenders like matrix or tensor factorization,
that can more easily incorporate negative feedback.

% To understand the effect of incorporating pair-level feedback on 
% diversity of the recommendations (Q6), 
% we plot the average pairwise cosine similarity of top-$10$ recommendation items of three different setups 
% for each user in 
% Fig.~\ref{fig:rec-diversity-dist}. As evident in the figure, compared to the baseline (Item-level), 
% setups that utilize pair-level feedback do
% not always reduce the diversity of the recommendations. 
% For instance, in the book domain, 
% after incorporating pair-level 
% feedback within Pair-level (Exp-5) setup, $11$ users 
% received more diverse recommendations among which $5$ 
% users experienced improvement of recommendation (measured by P@$10$).

% \begin{figure}[t]
%     \centering
%       \begin{subfigure}{.5\columnwidth}
%         \centering
%         \captionsetup{width=.9\columnwidth}
%         \includegraphics[width=\linewidth]{images/pairwise_sim.png}
%         \caption{Movies}
%         \label{fig:rec-diversity-dist-movielens}
%       \end{subfigure}%
%       \hfill
%       \begin{subfigure}{.5\columnwidth}
%         \centering
%         %\captionsetup{width=.8\columnwidth}
%         \includegraphics[width=\linewidth]{images/goodreads/pairwise_sim.png}
%         \caption{Books}
%         \label{fig:rec-diversity-dist-goodreads}
%       \end{subfigure}%
%       \vspace{-1\baselineskip}
%       \caption{Average pairwise cosine similarity of top-$10$ recommendations per user.}
% 	\label{fig:rec-diversity-dist}
% 	\vspace*{-0.5cm}
% \end{figure}

Finally, we show anecdotal examples from our user study in Fig.~\ref{fig:anecdotes}.
For movie recommendation (Fig.~\ref{fig:anecdote-movies}), 
incorporating user feedback on pairwise similarity introduces new items
into the top-$10$ recommendations (\textit{The Chronicles of Narnia}, \textit{The BFG})
for their respective users. These new recommendations possess the similarity aspects
liked by the user (\textit{fantasy} for \textit{Narnia}, \textit{based on a book} for \textit{The BFG}), and lack the dimensions that the user has implicitly disliked (\textit{crime} for the first anecdote, 
\textit{fiction} for the second one). Similarly, we present  
two instances of improvement in book recommendation in Fig.~\ref{fig:anecdote-books}, where 
incorporation of pair-level feedback results in reducing the relevance score of disliked 
items (\textit{Memoirs of a Geisha}, \textit{DotCom Secrets}) 
and bringing up more relevant items (\textit{The Iliad}, 
\textit{Mindset: The New Psychology of Success}) for the respective users in the ranked list of recommendations. 

% !TeX root = ../2021-www-fp-elixir.tex
% \noindent \textbf{Limitations imposed by resource constraints}
% \label{sec:discussion}

%(alternative frameworks, extensions, high level justifications,
%constrasting datasets, ...)
% other questions, whether the explnation is relevant. 
% what if we have ther types of actions like reviews (multi-model)
% other values for feedback maybe rating
% similarities might not exist , change the question?
\noindent \textbf{Limitations imposed by resource constraints.} One limitation of this work is the scale of the user study. 
Evaluating ELIXIR on a larger scale
% (e.g., 50-100 users)
% larger is subjective, from one order to magnitude to deployment-scale
would incur
% prohibitive
substantially more monetary cost and require 
design and implementation of a large-scale system suitable for 
orchestrating and monitoring the longitudinal process of user-system interactions.
Resource constraints also impact the possibility of full exploration of the parameter space in this work, 
such as a thorough search for the best number of latent dimensions $d$, as that might require the repetition of 
the whole study. Nevertheless, 
we evaluated another value 
% the latent dimension $d$
$d=10$
in the movie domain to verify the robustness of \elixir.
% w.r.t. number of features.
% already clear
Trends were very similar: P@$5$ values for item-level, pair-level (top explanations), 
and item+pair level (top explanations) 
came out to be $0.520$, $0.712$, and $0.712$, respectively,
retaining previously observed statistically significant trends of superiority of configurations involving pair-level feedback over item-level only.
% which again shows the significant improvement 
% achieved through incorporation of pair-level feedback. 
%other transformation functions 
%implicit pair-level feedback

	% \input{sections/05-discussion}
	% !TeX root = ../2021-www-fp-elixir.tex
\section{Related Work}
\label{sec:related}

\subsection{Explaining recommendations}
\label{subsec:exp-rec}

Explaining recommendations is a key component towards transparency, 
user satisfaction and scrutability
~\cite{tintarev2012evaluating, balog2020measuring}. 
Herlocker et al.~\cite{herlocker2000explaining} presented the first 
work on explainability for collaborative-filtering based recommenders
by locating people with similar profiles (neighbors). With the introduction of 
the Latent Factor Model (LFM) by Koren~\cite{koren2008factorization}, research on 
interpretability of latent features gained 
attention~\cite{zhang2014explicit, peake2018explanation, cheng2019incorporating}.
To mitigate the sparseness of ratings in standard collaborative filtering, 
hybrid recommenders based on 
Knowledge Graphs (KGs) were introduced. Explanations in these models 
are mostly path-based, that is, a recommendation item is explained by finding 
the most relevant paths from the user 
node~\cite{ai2018learning, wang2018ripplenet, xian2019reinforcement, wang2019explainable, musto2019linked}. 
Explainability is particularly important in sophisticated neural models, 
that 
% Explainable deep models
mostly use the attention mechanism 
over words~\cite{seo2017interpretable}, reviews~\cite{chen2018neural}, 
items~\cite{cheng2019incorporating} or images~\cite{chen2019personalized}
to learn their importance for a given recommendation. Another line 
of work focuses on generating post-hoc explanations for models 
that lack transparency~\cite{peake2018explanation, wang2018reinforcement, ghazimatin2020prince, musto2020generating}. 
For example, Ghazimatin et al. proposed 
a method for generating minimal and counterfactual explanations for 
recommenders with random walk at their core~\cite{ghazimatin2020prince}.
Yet another example is in LFMs where post-hoc explainability 
can be approached via using association rule mining~\cite{peake2018explanation}. 

\subsection{Critique-based recommenders}
\label{subsec:critique}

In most of the prior works on explainable recommendation, the role of explanations is 
limited to providing users with insights into the recommendation model. This limits 
scrutability as users might not have a clue as to how to correct the system's reasoning. 
To increase user control over the recommendation process, critique-based recommenders 
were introduced~\cite{chen2007hybrid, chen2012critiquing}. 
%and remain cosmetic by value with regard to improving the core recommender model.
Critiquing is a method for conversational (a.k.a. sequential and interactive) recommendation that adapts recommended items in response to
user preferences on item attributes.
% Providing explanations for recommended items not only allows users to understand the reason for 
% receiving recommendations but also provides users with an opportunity to refine recommendations
% by critiquing undesired parts of the explanation. 
Incremental critiquing/tuning~\cite{mccarthy2010experience,reilly2004incremental, lee2020explanation} was thus proposed to improve recommendation quality 
over successive recommendation cycles. 
However, this restricts users to critique/tune based on explicit item properties 
which are hard to generalize. 
Recent works in the form of Deep Language-based Critiquing
(DLC)~\cite{wu2019deep,luo2020latent} address this challenge by accepting arbitrary language-based critiques to improve the
recommendations for latent factor based recommendation models.
In~\cite{luo2020deep}, Luo et al. improve the complexity of the existing critiquing frameworks by revisiting 
critiquing from the perspective of Variational Autoencoder (VAE)-based recommendation methods and keyphrase-based interaction.
Existing critique-enabled recommenders mostly focus on 
negative feedback on concrete features of individual recommendation items. In \elixir, we address 
this limitation by enabling users to give both positive and negative feedback on pairs of recommendation and explanation items. 

\subsection{Set-based preference}
\label{subsec:set-pref}

Most recommendation approaches rely on signals provided by users on individual
items. Another mechanism of eliciting preference is to ask users for feedback on itemsets. 
Such set-based preference annotations help in faster learning of user interests,
especially in cold start situations~\cite{chang2015using}.
Moreover, users who are not willing to provide explicit feedback on individual items due 
to privacy concerns may agree to providing a single rating to a set of items, as it provides a certain level of information
abstraction. At the same time, from the given set-based rating, some information regarding item-wise preference can 
be inferred. In the same vein, in~\cite{sharma2019learning}, authors gathered users' preferences on itemsets and 
developed a collaborative filtering method to predict ratings for individual items in the set.
Apart from understanding user profiles, set-based learning is also 
useful in works that have focused on recommending lists of items or
bundles of items to users such as recommendation of music playlists~\cite{aizenberg2012build}, travel packages~\cite{liu2011personalized}, and
reading lists~\cite{liu2014recommending}. \elixir reinforces this viability of set-based feedback.

	% !TeX root = ../2021-www-fp-elixir.tex
\section{Conclusion}
\label{subsec:confut}

In this work, we have shown how explanations for recommendations can 
be made actionable by incorporating user feedback on pairs
of items 
%($rec$ and $exp$) 
into recommender systems. 
\elixir is a % applies the concept of
human-in-the-loop
% where the
system
that 
proactively elicits lightweight user feedback on
the similarity of recommendation and explanation pairs.
\elixir subsequently densifies this feedback 
using a smart combination of label propagation and locality sensitive hashing,
learns user-specific item representations by means of a soft-constraint-regularized
optimization, 
and seamlessly 
injects 
%pair-level signals 
these learned signals
into the underlying recommender.
% via user-specific item vectors.
We instantiated this framework with one major family of recommender models,
based on random walks with restart and exemplified by the \recwalk method.
Our experimental evaluation, based on a longitudinal user study,
showed major gains in recommendation quality. % and user satisfaction.
This demonstrates the power of the proposed \elixir framework
to learn more discriminative latent features about user preferences,
which are disregarded in traditional item-level ratings.
%As an instantiation of this more
%general idea, we show how one can exploit system-generated explanations beyond
%their role in building trust and satisfaction to something more actionable:
%making effective use of preference signals on
%the similarity of recommendation and explanation pairs.
%Our proposed framework, 
%
%\elixir operates in an active learning manner where the system
%proactively elicits light-weight feedback on
%the similarity of recommendation and explanation pairs.
%% that it deems to be the most worthwhile
%%for improving the underlying model.
%\elixir subsequently densifies this feedback 
%using a smart combination of label propagation and locality sensitive hashing,
%and injects pair-level signals 
%into the recommender via user-specific item vectors.

Future work would naturally focus on extending \elixir to other families
of recommenders such as matrix/tensor factorization or neural methods, 
exploring alternative strategies for absorbing
pairwise feedback, and investigating the effectiveness of \elixir
for long-tail users with sparse profiles.

	\section*{Acknowledgements}
	This work was supported by the ERC Synergy Grant 610150 (imPACT).
	%\clearpage
	\balance
	
	\bibliographystyle{ACM-Reference-Format}
	\bibliography{elixir}
	
\end{document}